\def\zem{$z_{\rm em}$}
\def\zabs{$z_{\rm abs}$}
\def\zphot{$z_{\rm phot}$}

\def\hi{H~{\sc i}}

\def\nhi{\mbox{$\sc N(\sc H~{\sc I})$}}
\def\lognhi{\mbox{$\log \sc N(\sc H~{\sc i})$}}

\def\caii{Ca~{\sc ii}}

\def\cii{C~{\sc ii}}
\def\civ{C~{\sc iv}}
\def\feii{Fe~{\sc ii}}

\def\mgi{Mg~{\sc i}}
\def\mgii{Mg~{\sc ii}}

\def\nii{[N~{\sc II}]}
\def\sii{S~{\sc i}}
\def\siii{Si~{\sc ii}}
\def\siiii{Si~{\sc iii}}
\def\siiv{Si~{\sc iv}}
\def\tiii{Ti~{\sc ii}}

\def\alii{Al~{\sc ii}}

\def\znii{Zn~{\sc ii}}
\def\crii{Cr~{\sc ii}}

\def\oi{O~{\sc i}}
\def\oii{[O~{\sc ii}]}
\def\oiii{[O~{\sc iii}]}
\def\ovi{O~{\sc vi}}
\def\ni{N~{\sc i}}
\def\sii{S~{\sc ii}}

\documentclass[useAMS,epsfig]{mn2e}

\usepackage{rotating}

\usepackage{graphicx}

\title[New SINFONI DLAs Detections]{A SINFONI Integral Field Spectroscopy Survey for Galaxy Counterparts to Damped Lyman-$\alpha$ Systems - I. New Detections and Limits for Intervening and Associated Absorbers. \thanks{Based on observations collected during programmes ESO 79.A-0673 and 80.A-0742 at the European Southern Observatory with SINFONI on the 8.2 m YEPUN telescope operated at the Paranal Observatory, Chile.} }
\author[C\'eline P\'eroux et al.] {C\'eline P\'eroux$^1$\thanks{e-mail:celine.peroux@gmail.com}, Nicolas Bouch\'e$^{2,3}$, Varsha P. Kulkarni$^4$,
Donald G. York$^5$,
\newauthor
 \& Giovanni Vladilo$^6$\\
$^1$ Laboratoire d'Astrophysique de Marseille, OAMP, Universit\'e Aix-Marseille \& CNRS,\\
38 rue Fr\'ed\'eric Joliot Curie, 13388 Marseille cedex 13, France  \\
$^2$ Max-Planck-Institut f\"ur extraterrestrische Physik Giessenbachstrasse, 85748 Garching, Germany \\
$^3$ Department of Physics, University of California, Santa Barbara, CA 93106, USA.\\
$^4$ Dept. of Physics and Astronomy, Univ. of South Carolina, Columbia, SC 29208, USA.\\
$^5$ Dept. of Astronomy and Astrophysics, Univ. of Chicago, 5640 S. Ellis Ave, Chicago, IL 60637, USA.\\
$^6$ Osservatorio Astronomico di Trieste - INAF, Via Tiepolo 11 34143 Trieste, Italy.
}
\begin{document}


\pagerange{\pageref{firstpage}--\pageref{lastpage}} \pubyear{2002}

\maketitle

\label{firstpage}

\begin{abstract}
Detailed studies of Damped and sub-Damped Lyman-$\alpha$ systems (DLA), the galaxies probed by the absorption they produce in the spectra of background quasars, rely on identifying the galaxy responsible for the absorber with more traditional methods. Integral field spectroscopy provides an efficient way of detecting faint galaxies near bright quasars, further providing immediate redshift confirmation. Here, we report the detection of H-$\alpha$ emission from a DLA and a sub-DLA galaxy among a sample of 6 intervening quasar absorbers targeted. We derive F(H-$\alpha$)=7.7$\pm$2.7$\times$10$^{-17}$ erg/s/cm$^2$ (SFR=1.8$\pm$0.6 M$_{\odot}$/yr) at impact parameter b=25 kpc towards quasar Q0302$-$223 for the DLA at \zabs=1.009 and F(H-$\alpha$)=17.1$\pm$6.0$\times$10$^{-17}$ erg/s/cm$^2$ (SFR=2.9$\pm$1.0 M$_{\odot}$/yr) at b=39 kpc towards Q1009$-$0026 for the sub-DLA at \zabs=0.887. These results are in line with low star formation rates previously reported in the literature for quasar absorbers. We use the \nii $\lambda$ 6585/H-$\alpha$ ratio to derive the HII emission metallicities and compare them with the neutral gas \hi\ absorption metallicities derived from high-resolution spectra. In one case, the absorption metallicity is actually found to be higher than the emission line metallicity. For the remaining objects, we achieve 3-$\sigma$ limiting fluxes of the order F(H-$\alpha$)$\sim$10$^{-17}$ erg/s/cm$^2$ (corresponding to SFR$\sim$ 0.1 M$_{\odot}$/yr at z$\sim$1 and $\sim$1 M$_{\odot}$/yr at z$\sim$2), i.e. among the lowest that have been possible with ground-based observations. We also present two other galaxies associated with \civ\ systems and serendipitously discovered in our data. 

\end{abstract}

\section{Introduction}

Observations of quasar absorbers allow a study of the gaseous component of the high-redshift galaxies (e.g., Bahcall et al. 1969). Indeed, the neutral hydrogen content of the strongest of these quasar absorbers, the so-called Damped Lyman-$\alpha$ systems (DLAs), systems with log \nhi$>$20.3 atoms/cm$^2$  has been measured in several hundred objects (Prochaska et al. 2005, Noterdaeme et al. 2009). These systems exhibit strong damping wings, however, technically any absorption system with a Doppler parameter  b $<$ 100 km/s  and  log \nhi$>$19 atoms/cm$^2$  will exhibit damping wings. These systems have been dubbed sub-Damped Lyman-$\alpha$ systems (sub-DLAs) by P\'eroux et al. (2003a) and have been studied extensively in recent year (e.g. P\'eroux et al. 2003b, P\'eroux et al. 2005). Nevertheless, studying the stellar content of all these systems turns out to be rather challenging: the galaxies that produce such absorption might be faint, thus requiring deep observations to detect their stellar/interstellar emission; in addition, they have small angular separations from the bright background quasars, which makes it difficult to disentangle the light of the galaxy from that of the quasar (e.g., Rao et al. 2003, Chen et al. 2003 and Rosenberg \& Schneider 2003). Deep observations have been made possible recently thanks to a hundred-hour long exposure at the VLT that revealed a population of extended Ly-$\alpha$ emitters which can be identified with the elusive host population of DLAs (Rauch et al. 2008). In trying to image the implied H-$\alpha$ from an absorption galaxy known from its absorption in a background quasar, there is a severe limitation due to small angular separation only partly overcome with the use of space-based Hubble Telescope observations that provide higher spatial resolution (e.g. Kulkarni et al. 2000, Kulkarni et al. 2001, Bouch\'e et al. 2001, Bowen et al. 2001 and Weatherley et al. 2005).  There have also been some successful detections of candidate absorber galaxies using adaptive optics broad-band imaging (e.g., Chun et al. 2006 and Chun et al. 2010). But the complex nature of the adaptive optics point spread function limits the use of this method at small angular separations, corresponding to an impact parameter in the order of several kiloparsecs at the typical redshift of the absorption systems. More importantly, any broad-band identification of a candidate absorber galaxy (e.g. Fumagalli et al. 2010) requires follow-up spectroscopy in order to confirm that the emission redshift of the object corresponds to the absorption redshift measured in the quasar spectrum. This time-consuming step is sometimes missing, thus complicating the interpretation of the observations made so-far. As a result of these issues, only 15 spectroscopically confirmed identifications of sub-/DLA galaxies with measured \nhi\ are known at cosmological redshift z $<$ 1 (Rao et al. 2003, Chen et al. 2005 and references therein) and 3 at z $>$ 1 (e.g. Weatherley et al. 2005).

Imaging spectroscopy at near infra-red wavelengths with adaptive optics can be used to efficiently solve the above mentioned problems. The advantages afforded by this technique are manifold: 1) the contribution from the quasar is deconvolved from the absorber in spectral space thus allowing one to reach virtually null impact parameters if the flux of the galaxy is greater than the quasar continuum flux; 2) the data provide a spectrum of the absorbing galaxy for immediate redshift confirmation; 3) the detection of nebular emission lines such as H-$\alpha$, a robust estimator of star formation rates (Moustakas et al. 2006) redshifted to the near infra-red at z$>$0.6, can give clues to the stellar content of the galaxy and metallicity of the warm gas producing the nebular lines and 4) the spatially resolved kinematics data provide information on the dynamical state of the system (see Paper II of the series, P\'eroux et al. 2010).

The approach of using an integral field unit was already taken by Yanny et al. (1990), but most notably, the powerful combination of the 8-m class telescope and SINFONI instrument has been successfully pioneered by Bouch\'e et al. (2007). These authors have detected 67\%\ (14 out of 21) of the systems in a sample of \mgii\ absorbers at z$\sim$1. The H-$\alpha$ detections are used to derive star formation rate of the order 1 to 20  M$_{\odot}$/yr. Their observations are sensitive to fluxes F(H-$\alpha$)$>$1.2$\times$10$^{-17}$ erg/s/cm$^2$, corresponding to a star formation rates of $\sim$ 0.5 M$_{\odot}$ yr$^{-1}$. Interestingly, lower detection rates are found at higher redshifts around z$\sim$2 (Bouch\'e et al., private communication). 

Here, we report the first results of our \hi-selected survey at redshifts z$\sim$1 and z$\sim$2 aimed at detecting the host galaxies via their H-$\alpha$ signature. In the present work, we report the detections of 2 DLAs/sub-DLAs at z$\sim$1 for which \nhi\ and hence absorption metallicities are known from high-resolution spectroscopy out of the 5 quasar fields with intervening absorbers (one with two sub-DLAs) searched for. We also present two other galaxies associated with \civ\ systems and serendipitously discovered in our data. In this paper, we concentrate on the description of the data and other basic physical properties of the detections. In paper II (P\'eroux et al. 2010), we present the dynamical properties of these two detections. 

The present paper is structured as follows. Details of observations and data reduction processes are provided in Section 2. 
In the third section, we present the primary targets of our survey for which physical properties are presented in Section 4. Section 5 presents a description, for each field under study, of the non-prime a posteriori detections. 
Readers who are primarily interested in the physical properties of the primary targets may
skip directly to Sections 4, which are largely self-contained. In all the remaining, we assumed a cosmology with H$_0$=71 km/s/Mpc, $\Omega_M$=0.27 and $\Omega_{\rm \lambda}$=0.73.

\section{Observations and Data Reduction}

\begin{table*}
\begin{center}
\caption{Journal of Observations. }
\label{t:JoO}
\begin{tabular}{cccccccccc}
\hline\hline
Quasar 		  &Coordinates$^{\dagger}$ &Mag &\zem &\zabs &Observation Date &T$_{\rm exp} [sec]$ &Band &AO &seeing["]\\
\hline
Q0134$+$0051$^{a}$ &SDSS J013405.76+005109.3	&18.4	&1.522	&0.842	&27-29/09/07	&16$\times$600 &J &no AO	&0.8\\
Q0302$-$223$^b$	    &03 04 49.86 -22 11 51.9                &16.0	&1.409	&1.009	&28/10, 23/12/07		&4$\times$600$+$8$\times$600 &J   & NGS &0.7\\
Q0405$-$331$^{c,d}$  &04 07 33.91 -33 03 46.4 	  &19.4	&2.562	&2.569	&14/10, 01/11/07	&10$\times$600$+$4$\times$520 &K  & NGS &0.6\\
Q1009$-$0026$^e$	    &10 09 30.47 -00 26 19.1	           &17.5	&1.242	&0.843	&06/12/07,03/01/08 &5$\times$600$+$4$\times$900 &J  &no AO &1.0\\
Q1009$-$0026$^e$	    &--                                                          &--	         &--	         &0.887		     &-- &-- &-- &-- &--\\
Q1228$-$113$^d$	    &12 30 55.62 -11 39 09.9                 &22.0  	&3.528	&2.193	&06-16/01,13/02/08 &16$\times$600$+$4$\times$520 &K  & NGS&0.4\\
Q1323$-$0021$^f$	    &SDSS J132323.78-002155.2    &18.2	&1.388	&0.716	&07/04, 23/05, 12/07/07	&13$\times$900 &J		&no AO &0.8\\
\hline\hline 				       			 	 
\end{tabular}			       			 	 
\end{center}			       			 	 
\begin{minipage}{140mm}
{\bf Note:} An 8"$\times$8" field of view was used, corresponding to a spaxel size of 0.25". Magnitudes are V magnitudes except for Q0405$-$331 and Q1228$-$113, where there are B magnitudes.\\
{\bf no AO:} no Adaptive Optics, natural seeing.\\
{\bf NGS:} Adaptive Optics with a Natural Guide Star.\\
$^{\dagger}$ SIMBAD coordinates unless the quasar is part of SDSS, in which case SDSS names are provided.\\
{\bf $^a$:} P\'eroux et al. 2006b.\\
{\bf $^b$:} Pettini et al. 2000.\\
{\bf $^c$:} the DLA in this quasar has \zabs$\sim$\zem.\\ 
{\bf $^d$:} Akerman et al. 2005.\\
{\bf $^e$:} Rao et al. 2006, Meiring et al. 2007.\\
{\bf $^f$:} P\'eroux et al. 2006a.\\
\end{minipage}
\end{table*}

{\bf Target Selection:} We have targeted the H-$\alpha$ emission of 6 intervening and one associated quasar absorbers with \hi\ column densities log \nhi$>19.0$ atoms cm$^{-2}$, corresponding to the definition of sub-DLAs (P\'eroux et al. 2003a). 
All of the absorbers have well-constrained \nhi\ based on space-based HST observations (Rao, Turnshek \& Nestor 2006). Similarly, the objects have been selected so that the metallicity of each of these systems has been studied with high-resolution spectroscopy. The systems were selected to have metallicities free from dust-depletion, as traced by zinc, to be around or greater of a tenth solar, [Zn/H]$>$-1.0\footnote{We use the usual definition [X/H]=log[X/H]-log[X/H]$_{\odot}$}. Because the redshift of the absorbers is accurately known from metal lines fits, we have also selected absorbers for which the predicted H-$\alpha$ lines fall in regions free from OH sky lines contamination. This resulted in a sample of absorbers, the so-called primary targets, with absorption redshift 0.7 $<$ \zabs $<$ 2.5, shifting the H-$\alpha$ emission line to SINFONI\footnote{See the following url for more information: http://www.eso.org/sci/facilities/paranal/instruments/sinfoni/}  $J$ or $K$-band filters. Note that the flux sensitivity is enhanced in the $K$-band compared to that in the $J$-band. There is a loss in raw instrument throughput in the $J$-band with respect to $K$-band, which is more than compensated by the fact that we are observing lower redshift targets in the $J$-band with respect to the $K$-band. A journal of observations summarising the absorber redshift, \nhi\ column density and metallicity with appropriate reference for each of the systems is presented in Table~\ref{t:JoO}. 

\noindent
{\bf Observations:} The observations were carried out during three separate observing runs (ESO 79.A-0673, 80.A-0330 and 80.A-0742) at the European Southern Observatory with SINFONI on the 8.2 m YEPUN telescope. 
Table~\ref{t:JoO} provides the date of observations, the exposure times for each of the objects and the resulting seeing. For fields where the quasar itself is bright enough or for which a bright star is available nearby, we have used it as a natural guide star (NGS) for adaptive optics (AO)  in order to improve the spatial resolution. For the fields with no suitable tip-tilt stars, we have used no AO (natural seeing). The resulting seeing ranges from 0.4 to 1.0". For each quasar field, we took  several 520 to 900 sec exposures, placing the quasar in the four quadrants of the 8$\times$8" field of view. This allows for both a robust sky subtraction and an increased search radius. The total exposure time at the position where the absorbing galaxy is searched for thus varies depending on its location with respect to the quasar, it is higher at the center of the field, where the quasar could potentially contaminate the detections. The $J$ ($K$) grism provides a spectral resolution of around R$\sim$2000 (4000). 

\noindent
{\bf Data Reduction:} Two of us (NB and CP) independently reduced the data for one of the quasar fields with the pipeline developed by MPE on one hand (SPRED, Abutter et al. 2006, F\"orster-Schreiber et al. 2009) and with the publicly available pipeline provided by ESO (version 1.9.8, see: http://www.eso.org/sci/data-processing/software/pipelines/sinfoni/sinfo-pipe-recipes.html) on the other hand. After some adjustments to optimise the ESO-pipeline input parameters, the results were found to be consistent with each other and all the other data were reduced using the ESO pipeline together with custom codes for correction of detector bad columns, cosmic rays removal, OH line suppression and sky subtraction (Davies et al. 2007) and flux calibration. 

The main steps of the procedure are as follows. Master bias and flat images were constructed using calibration cubes taken closest in
time to the science frames and used to correct each data cube. The science frames were pair-subtracted with an ON-OFF pattern to eliminate variation in the infra-red sky background. The wavelength calibration is based on the Ar lamp. For each set of observations, a flux standard star was observed at approximately the same time and airmass and was reduced in the same way as the science data. These flux standard stars were then used for flux calibration by fitting a black body spectrum to the O/B stars or a power law to the cold stars and normalising them to the 2MASS magnitudes. These spectra were also used to remove atmospheric absorption features from the science cubes. The different observations were then combined spatially using the position of the quasar in each frame.

The analysis was performed with the QFitsView (http://www.mpe.mpg.de/$\sim$ott/dpuser/qfitsview.html) software. This 3D visualisation tool allows one to interactively select the area around the expected H-$\alpha$ emission line in wavelength space and subtract the quasar continuum using a small area around the line. QFitsView was mostly used for visual inspection and confirmation of the sources.

\section{Search for Primary Targets}

Out of the 5 useful quasar fields (one with 2 absorbers), we report 2 secure detections of H-$\alpha$ emission lines at the expected positions. Custom codes are used to extract the fluxes around a radius of 13 pixels for the DLA towards Q0302$-$223 and 7 pixels (due to edge effects) for the sub-DLA towards Q1009$-$0026. The fluxes are estimated from a Gaussian fit to the emission lines. Using the H-$\alpha$ luminosity, we then derive the star formation rate (SFR) assuming the Kennicutt (1998) flux conversion corrected to a Chabrier (2003) IMF. For all the absorbers searched for, we reach stringent star formation limits. The 5-$\sigma$ upper limits for non-detections are computed for an unresolved source spread over 32 spatial pixels and spectral FWHM = 6 pixels = 9 \AA. No limits are derived for the absorber towards Q0405$-$331 because of the presence of the emission line from the quasar itself preventing us from putting stringent limits at small radius. These results are described in detail the following section. 

\begin{table*}
\begin{center}
\caption{Summary of the detections and upper limits for the primary targets.}
\label{t:results}
\begin{tabular}{ccccccccc}
\hline\hline
Quasar 		  &\zabs &\lognhi &EW(\mgii\ $\lambda$ 2796)&[Zn/H]  &$\Delta v^a$&F(H-$\alpha$) &Lum(H-$\alpha$) &SFR  \\
 		    &&[atoms/cm$^2]$ &[\AA]& &[km/s]&[erg/s/cm$^2]$ &[erg/s]&[M$_{\odot}$/yr]  \\
\hline
Q0134$+$0051{\bf $^b$}      &0.842 &19.93$^{+0.10}_{-0.15}$  &1.15$\pm$0.01$^c$	&$<-$0.36		     &160 &$<$1.0 $\times$10$^{-17}$		&$<$0.3$\times$10$^{41}$	    &$<$0.2 \\
Q0302$-$223			     &1.009 &20.36$^{+0.11}_{-0.11}$  &1.16$\pm$0.04$^d$	&$-$0.51$\pm$0.12     &-- &7.7$\pm$2.7$\times$10$^{-17}$&4.1$\pm$1.4$\times$10$^{41}$&1.8$\pm$0.6 \\
Q1009$-$0026{\bf $^b$}	     &0.843 &20.20 $^{+0.05}_{-0.06}$ &0.713$\pm$0.038$^d$	&$<-$0.98		     &60 &$<$1.0$\times$10$^{-17}$		 &$<$0.3$\times$10$^{41}$	    &$<$0.2	\\
--			     		     &0.887	&19.48  $^{+0.05}_{-0.06}$ &1.900$\pm$0.039$^d$	&$+$0.25$\pm$0.06     &300 &17.1$\pm$6.0$\times$10$^{-17}$ &6.6$\pm$2.3$\times$10$^{41}$&2.9$\pm$1.0\\
Q1228$-$113			     &2.193 &20.60$^{+0.60}_{-0.60}$   &--  &$<-$0.22       	    &--  &$<$0.5$\times$10$^{-17}$		 &$<$2.0$\times$10$^{41}$	    &$<$0.9\\
Q1323$-$0021{\bf $^b$}	     &0.716 &20.21$^{+0.21}_{-0.18}$   &2.229$\pm$0.071$^d$ &$+$0.61$\pm$0.20 &215 &$<$1.0$\times$10$^{-17}$		 &$<$0.2$\times$10$^{41}$	    &$<$0.1\\
\hline\hline 				       			 	 
\end{tabular}			       			 	 
\end{center}			       			 	 
\vspace{0.2cm}
\begin{minipage}{140mm}
{\bf $^a$:} $\Delta v$ is the velocity spread over which the metal lines have been fitted.\\
{\bf $^b$:} The 5-$\sigma$ upper limits for non-detections are computed for an unresolved source spread over 32 spatial pixels and spectral FWHM = 6 pixels = 9 \AA. No limits are derived for the absorber towards Q0405$-$331 because of the presence of the emission line from the quasar itself.\\
{\bf $^c$:} P\'eroux et al. 2006b.\\
{\bf $^d$:} Rao, Turnshek \& Nestor, 2006.\\  
\end{minipage}
\end{table*}

\subsection{Q0134$+$0051, \zabs=0.842}

This quasar was first discovered by the Sloan Digital Sky Survey (SDSS, York et al. 2000, Schneider et al. 2002).
A \zabs=0.842 sub-DLA has been studied by Rao et al. (private communication) from their HST Space
Telescope Imaging Spectrograph (STIS) follow-up of the object. P\'eroux et al. (2006b) have published an \hi\ column density of 
log \nhi = 19.93$^{+0.10}_{-0.15}$. These authors also report the metallicity of the system derived by fitting 13 components over about 160 km/s to the high-resolution UV spectrum of the quasar. They obtain the following abundances with respect to solar: [Mg/H]$> -$1.44; [Fe/H]=$ -$0.91$\pm$0.16; [Zn/H]$< -$0.36;  [Cr/H]$<-$0.75 and [Mn/H]$<-$0.96. This absorber was the primary target of our H-$\alpha$ search but no lines were detected down to the limit specified in Table~\ref{t:results}. The corresponding limits on luminosity and star formation rate are also provided.

\subsection{Q0302$-$223, \zabs=1.009}

\begin{figure*}
\begin{center}
\includegraphics[height=9cm, width=10cm, angle=0]{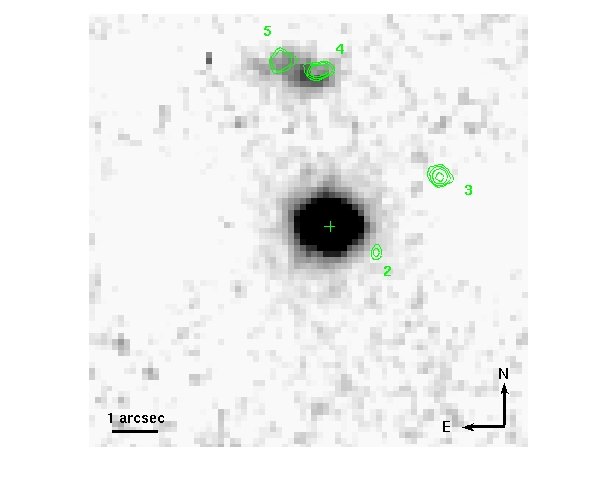}
\caption{SINFONI observations of the field of Q0302$-$223 spectrally collapsed at the position of the absorber at \zabs=1.009 - hence not showing the other galaxy at z=1.408 (Le Brun's object \#2). Object \#3 is not detected in the SINFONI data. The strong emission at the center is from the quasar itself. The green contours are from the HST/WFPC2 image of Le Brun et al. (1997) and clearly show that the objects \#4 and \#5 are two separate components. The numbering scheme is following Le Brun's original nomenclature. The green cross marks the position of the quasar in the HST/WFPC2 image. }
\label{f:Q0302_HST}
\end{center}
\end{figure*}

The absorber at \zabs=1.009 was first reported as a \mgii\ system by Petitjean \& Bergeron (1990), who also detected
\feii\ $\lambda$$\lambda$ 2586, 2600 and \mgi\ $\lambda$ 2852.  In this system, the \mgii\ absorption is a triple complex spanning
170 km/s. The \hi\ line was observed by Rao, Turnshek \& Briggs (1995) with the International UV Explorer (IUE).
A subsequent UV spectrum secured with the Faint
Object Spectrograph (FOS) on the Hubble Space Telescope
(HST ) confirmed that log \nhi\ 
= 20.33 (Pettini \& Bowen 1997, Boiss\'e et al. 1998, Rao \& Turnshek 2000). In addition, the FOS
spectrum reveals several strong features from \cii, \civ, \ni,
\oi, \siii, \siiii\ and \siiv. Pettini \& Bowen (1997) used William Herschel Telescope (WHT) observations to study \znii\
and \crii\ lines, which were then followed by observations with the HIRES spectrograph at Keck (Pettini et al. 2000). These authors observe that two main groups of components, separated by 36
km/s, produce most of the absorption seen in this DLA.
Additional weaker components, at v=35 and 121
km/s relative to \zabs=1.00945, are visible in the stronger
\feii\ lines. The resulting abundances with respect to solar are: [Zn/H]=$-$0.51$\pm$0.12, [Si/H]=$-$0.73$\pm$0.12, [Mn/H]=$-$1.32$\pm$0.12,~[Cr/H]=$-$0.96$\pm$0.12, [Fe/H]=$-$1.20$\pm$0.12~and [Ni/H]=$-$1.04$\pm$0.12. Finally, Boiss\'e et al. (1998) derive an upper limit on molecular hydrogen N(H$_2$), which translates into a molecular fraction f(H$_2$) $<$ 4.0 $\times$ 10$^{-3}$ (where f(H$_2$) = 2N(H$_2$)/[(2N(H$_2$) + N(HI)]).

Le Brun et al. (1997) obtained HST/WFPC2 imaging of the 10" square field centered on the quasar Q0302$-$223. They found four
faint objects at impact parameters less than 5" (objects \#2 to \#5 in their nomenclature),
the closest being detected only after profile subtraction of the quasar. These are reproduced as green contours overlaid on the SINFONI collapsed image in Figure~\ref{f:Q0302_HST}. Le Brun et al. (1997) note that there is no strong emission line at the wavelength
of the expected \oii\ $\lambda$ 3727 line at z = 1.0095 in the spectra
of objects \#4 and \#5, unresolved in ground-based observations
(see Guillemin \& Bergeron 1997).
Consequently, they argue that objects \#2 and \#3 are the most likely damped Lyman-$\alpha$ absorber candidates. 
Chen \& Lanzetta (2003) have further obtained photometric redshifts for seven objects
at $\Delta\theta$ $<$ 15" from the quasar. These authors were unable
to obtain reliable photometric redshifts for objects \#2 and \#3 (in Le Brun's nomenclature).
Indeed, the two objects are well resolved in the space-based images
but are blended in the PSF of the quasar in the ground-based
images. In addition, objects \#4 and \#5 are blended
with each other in Chen \& Lanzetta's data, although they are well resolved
 in the HST images. Chen \& Lanzetta (2003) determined the best-fit
photometric redshift for the two objects combined on the
basis of the summed fluxes. The results indicated a consistent photometric
redshift estimate \zphot=0.96 for object \#4 with or
without including the second component \#5.

\begin{figure*}
\begin{center}
\includegraphics[height=7.cm, width=10cm, angle=0]{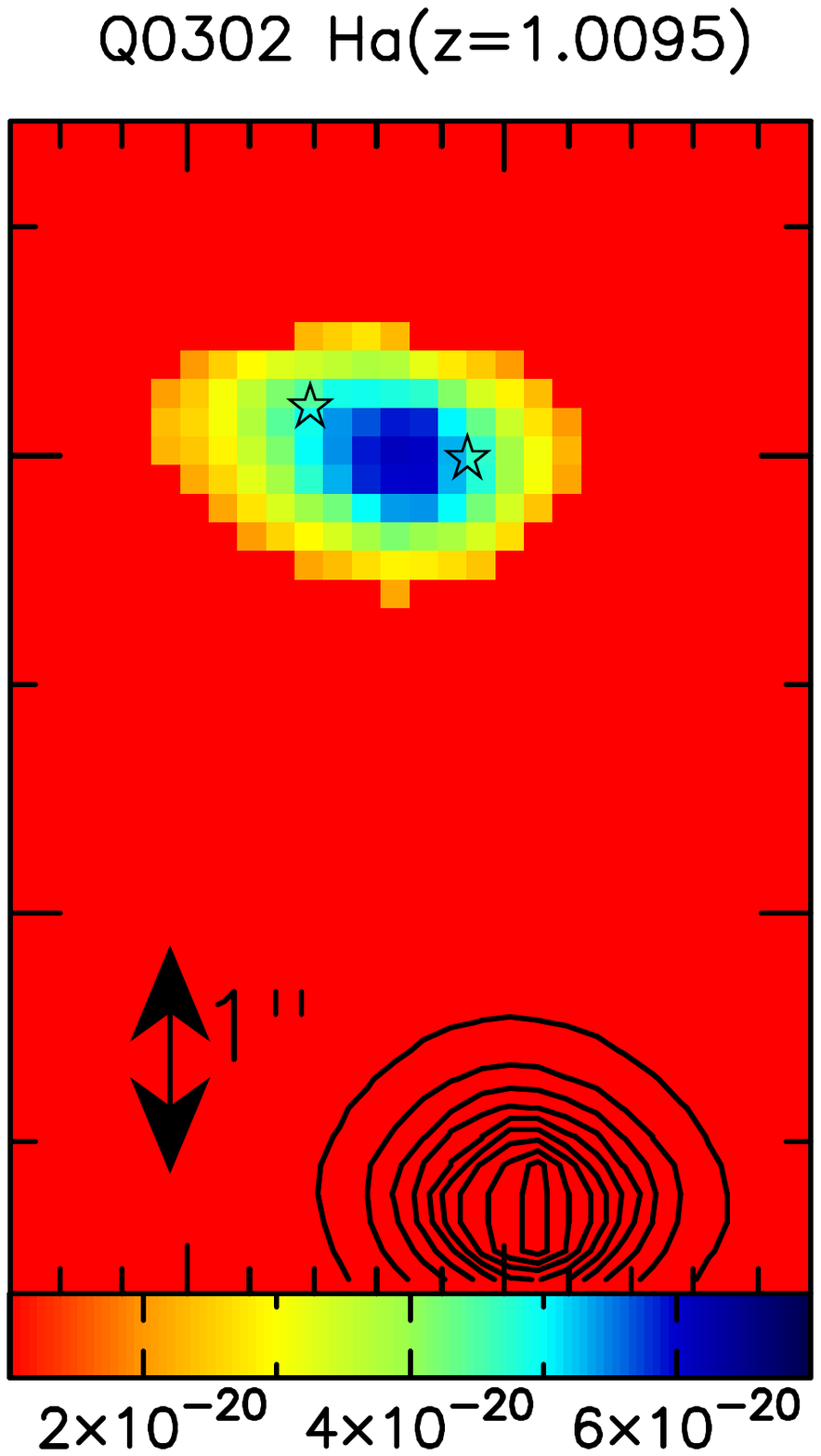}
\includegraphics[height=5cm, width=7.5cm, angle=0]{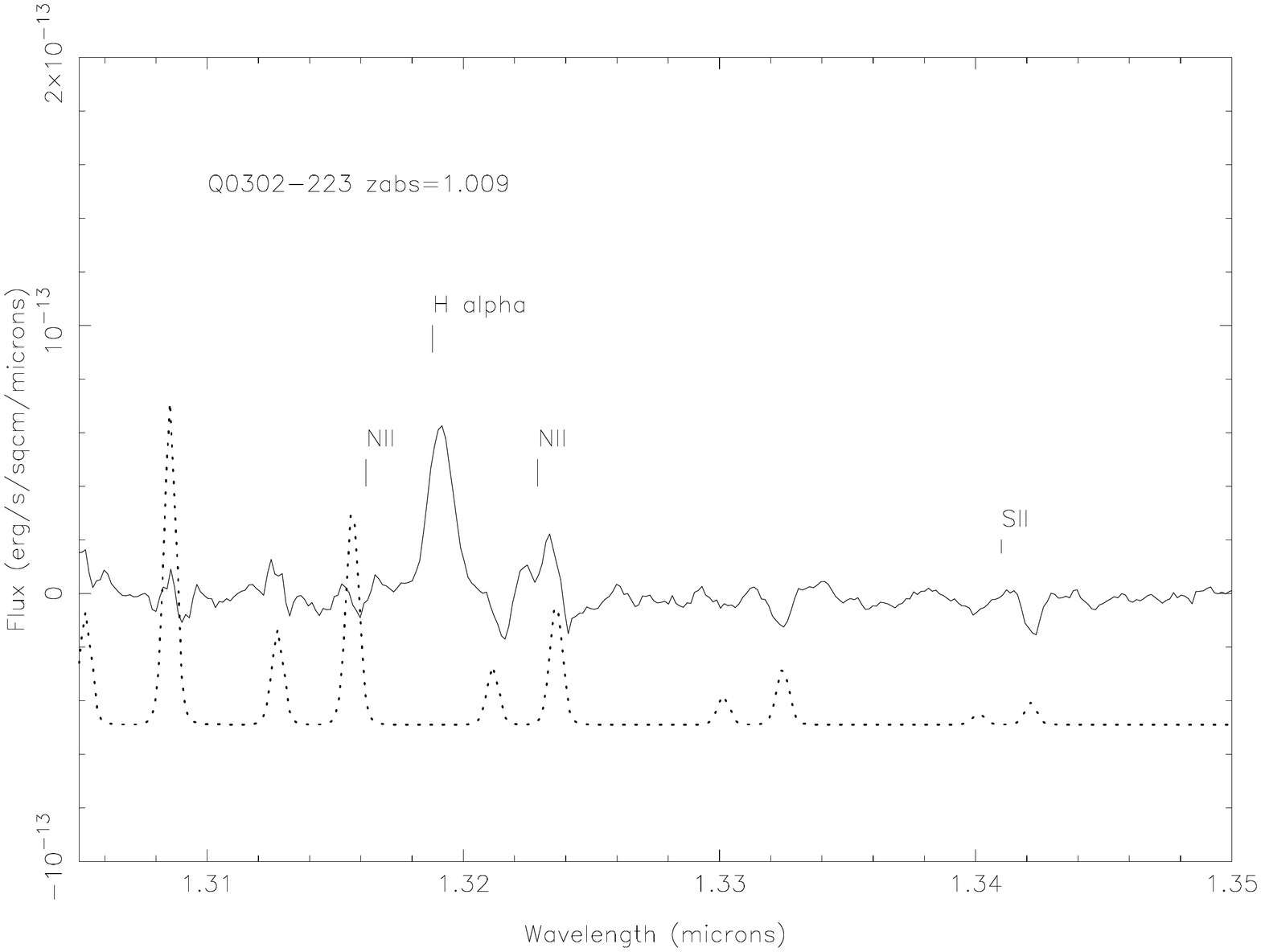}
\caption{{\bf Left Panel:} The H-$\alpha$ emission map of the targetted DLAin the field of Q0302$-$223. This galaxy corresponds to the blend of object \#4-5 detected in HST/WFPC2 images in Figure~\ref{f:Q0302_HST} by Le Brun et al. 1997. The thin lined, black contours indicate the position of the quasar. In this, and the following maps, north is up and east is to the left. The units are in erg/s/cm$^2$. At this redshift, 1"=8.1 kpc. {\bf Right Panel:} The integrated spectrum of the corresponding DLA galaxy with the expected H-$\alpha$, \nii\ $\lambda$ 6585 and \sii\ $\lambda$ doublet positions indicated at \zem=1.00946. The spectrum is smoothed (5 pixel boxcar). The dotted spectrum at the bottom of the panel is the sky spectrum indicating the position of the OH sky lines. In this case, \sii\ is not detected and an upper limit of \nii$<$2.6 $\times$ 10$^{-17}$ erg/s/cm$^2$ is derived due to contamination by an OH sky line.   }
\label{f:Q0302_Abs_spec}
\end{center}
\end{figure*}

In the SINFONI data presented here, objects \#4-5 are found to be associated with the DLA as suggested by the results from the photometric analysis of Chen \& Lanzetta (2003). The spectrum of the object shows clear emission from the H-$\alpha$ line as illustrated in Figure~\ref{f:Q0302_Abs_spec} which presents the integrated flux on an interval roughly twice the emission line's FWHM. Incidentally, object \#2 turns out to be an unrelated object associated with the quasar at \zem=1.40826 for which we detect \oiii\ emission in our SINFONI cube (see section 5 for further details). These findings serve as a caution to studies that look for galaxies responsible for quasar absorber solely based on imaging: there are cases where the closest object to the quasar line-of-sight is not the absorber and indeed the emitting galaxy is situated further away.

\subsection{Q0405$-$331, \zabs=2.569}

This quasar was part of the Parkes radio survey and its absorber was first reported by Ellison et al. (2001) as part of the CORALS survey which aimed at looking for absorbers in radio-selected quasars. These authors derived a column density of log \nhi=20.60. In this case the absorption redshift is similar to the emission redshift, making this absorber one of the so-called proximate DLA (PDLA; Ellison et al. 2002). Such a situation makes a search for galaxy responsible for the quasar absorber with IFU sub-optimal, since the emission line of the quasar is not kinematically decoupled from the absorbing-galaxy. 

This system is not detected in our data and no flux limits are derived because of the presence of the emission line from the quasar itself.

\subsection{Q1009$-$0026, \zabs=0.843 \& 0.887}

\begin{figure*}
\begin{center}
\includegraphics[height=7cm, width=9.cm, angle=0]{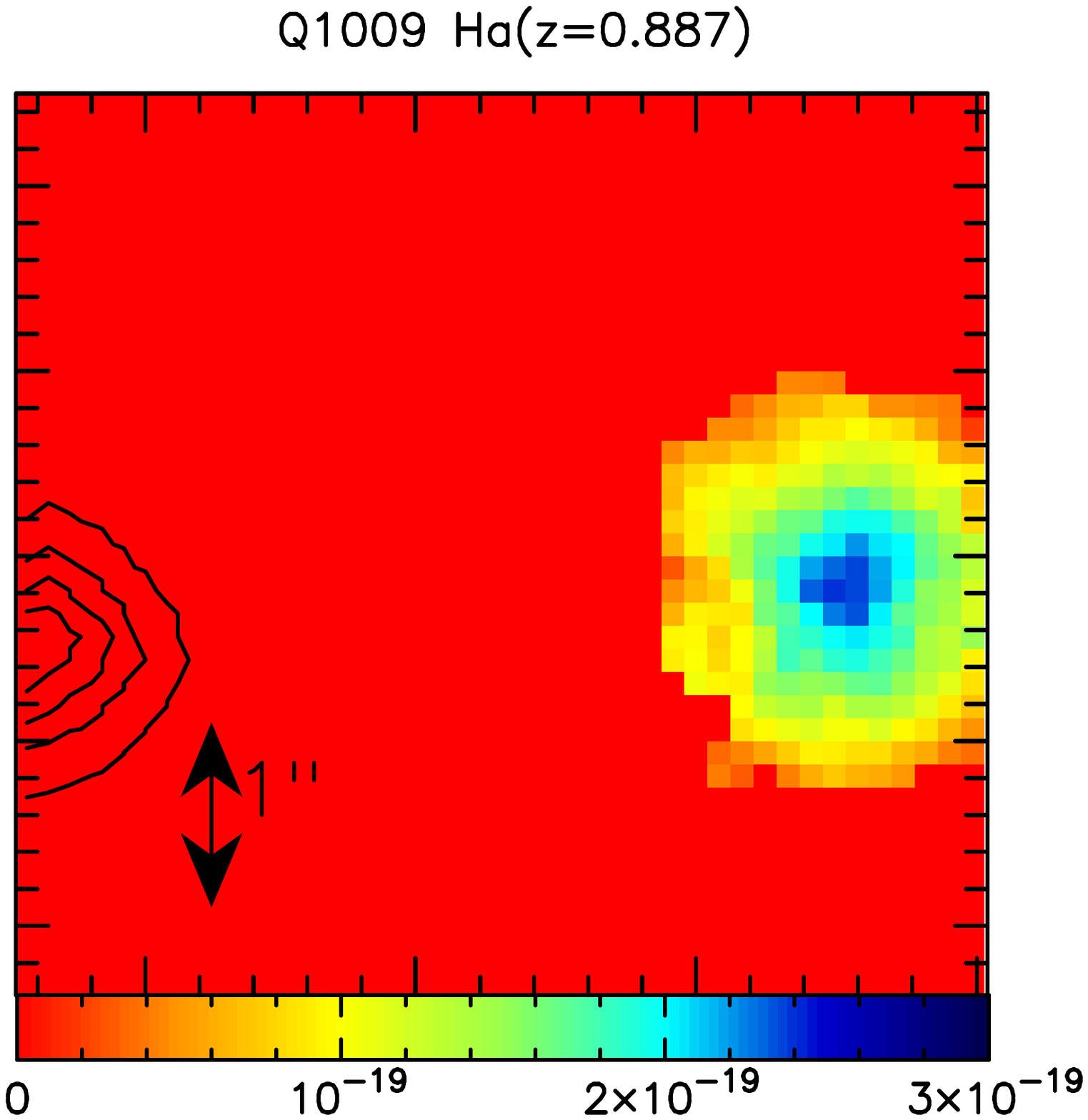}
\includegraphics[height=6cm, width=8.5cm, angle=0]{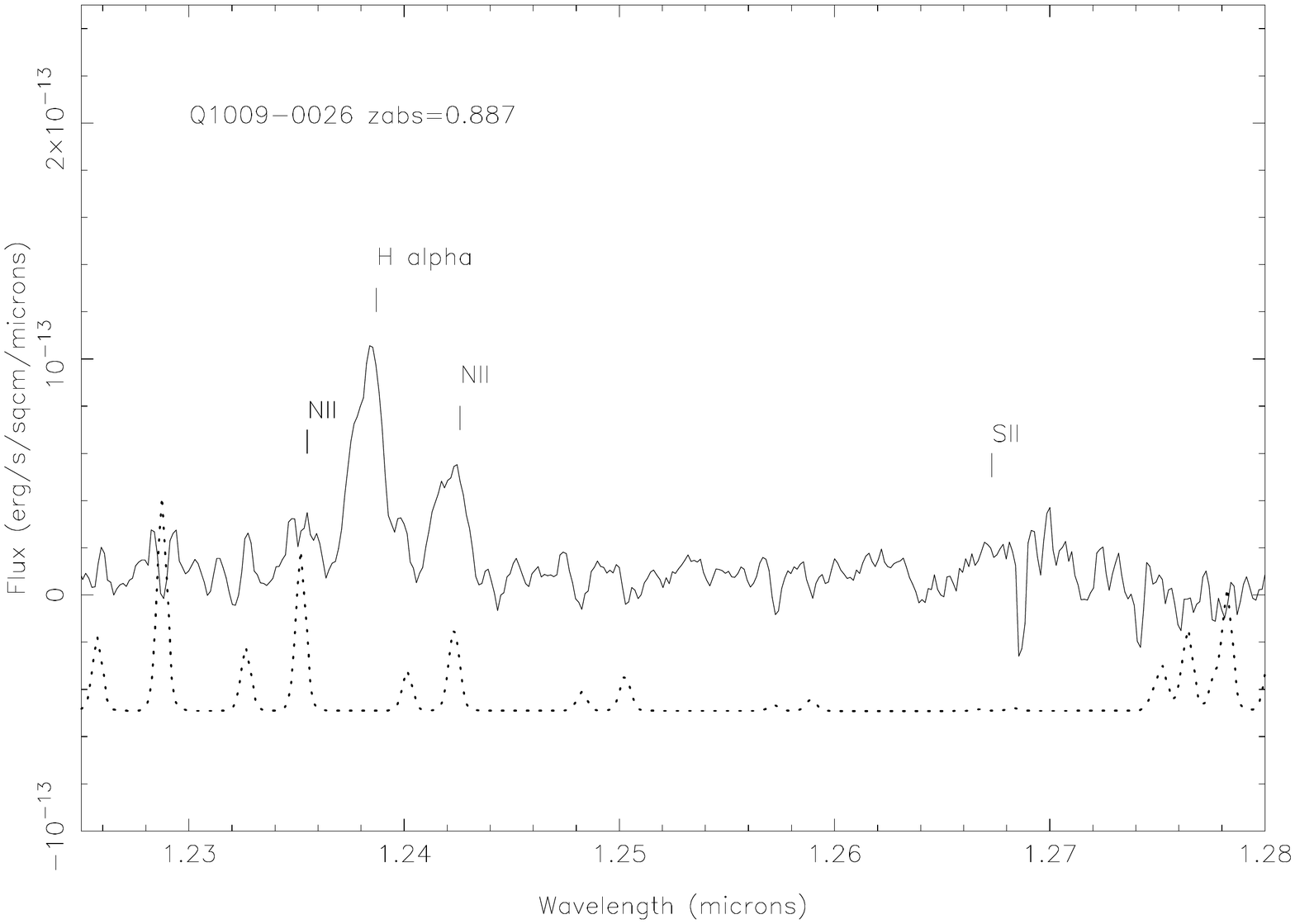}
\caption{{\bf Left Panel:} The H-$\alpha$ emission map of the targetted sub-DLA in the field of Q1009$-$0026. The thin lined, black contours indicate the position of the quasar. At this redshift, 1"=7.7 kpc.{\bf Right Panel:} The integrated spectrum of the galaxy with the expected H-$\alpha$, \nii\ and \sii\ doublet positions indicated at \zem=0.88637. The spectrum is smoothed (5 pixel boxcar).The dotted spectrum at the bottom of the panel is the sky spectrum indicating the position of the OH sky lines. \nii\ $\lambda$ 6585.27 is affected by an OH sky line but this is much narrower that the line we detect. We report F(\nii)$=$8.2$\pm$3.3 $\times$10$^{-17}$ erg/s/cm$^2$ which, given the rms of the spectrum, is consistent with the non-detection of \nii\ $\lambda$ 6549.86 assuming a ratio \nii\ $\lambda$ 6549.86/\nii\ $\lambda$ 6585.27=3 (Storey \& Zeippen 2000). \sii\ is not detected.  }
\label{f:Q1009_Abs_spec}
\end{center}
\end{figure*}

This is another quasar discovered as part of the SDSS survey. Rao, Turnshek \& Nestor (2006) reported two sub-DLAs along this line-of-sight: one at \zabs=0.8426 with log \nhi=20.20$^{+0.05}_{-0.06}$ and another one at \zabs=0.8866 with log \nhi=19.48$^{+0.01}_{-0.08}$. 

Meiring et al. (2007) have studied the metallicity of these systems using MIKE/Magellan data.  The system at \zabs=0.8426 shows a relatively
simple velocity structure extended over $\sim$60 km/s with only four components needed
to fit the observed profile. The resulting abundances with respect to solar are [Zn/H] $<-$0.98 from a non-detection and [Fe/H]=$-$1.28$\pm$0.07. The abundance ratios are therefore: [Fe/Zn]$>-$0.31, [Cr/Fe]$<-$0.16 and [Mn/Fe]=$-$0.16$\pm$0.06. This absorber is not detected in our SINFONI data.

The system at \zabs=0.8866 is fitted with seven components over more than 300 km/s. Meiring et al. (2007) were able to constrain the ionisation parameter for this
system using photo-ionisation CLOUDY modelling to be logU=-3.70 based on the observed Al III to Fe II ratio. The
\caii\ $\lambda\lambda$ 3933, 3969 lines were also detected in this system with log N(\caii)= 12.26. The abundances were found to be: [Zn/H]=+0.25$\pm$0.06 and [Fe/H]=$-$0.37$\pm$0.06. The abundance ratios are [Fe/Zn]=$-$0.63$\pm$0.06, [Si/Fe]$<-$0.39, [Ca/Fe]=$-$1.19$\pm$0.06, [Cr/Fe]$<-$0.65 and [Mn/Fe]$<-$1.24. Meiring et al. (submitted) have acquired broad-band photometry of this field with the Soar Optical Imager on the SOAR telescope using Sloan filters and detect an object possibly at the redshift of this absorber. Indeed, the same object is clearly detected in our SINFONI data. Figure~\ref{f:Q1009_Abs_spec} shows the H-$\alpha$ emission map and the spectrum of this galaxy with a strong H-$\alpha$ emission line. \nii\ $\lambda$ 6585.27 is affected by an OH sky line but this is much narrower that the line we detect. We report F(\nii)$=$8.2$\pm$3.3 $\times$10$^{-17}$ erg/s/cm$^2$ which, given the rms of the spectrum, is consistent with the non-detection of \nii\ $\lambda$ 6549.86 assuming a ratio \nii\ $\lambda$ 6549.86/\nii\ $\lambda$ 6585.27=3 (Storey \& Zeippen 2000). \sii\ is not detected.

It is interesting to note that the higher \nhi\ system (\zabs=0.843) is the most metal-poor of the two absorbers. It has the highest depletion of Fe compared to Zn (generally regarded as a higher depletion of Fe onto dust grains) and, by virtue of its non-detection by us, appears to have a lower SFR than the other sub-DLA detected in emission (\zabs=0.887).

\subsection{Q1228$-$113, \zabs=2.193}

Similarly to Q0405$-$331, this quasar was part of the Parkes radio survey and followed-up at optical wavelengths by Ellison et al. (2001) who first reported the DLA. These authors derived a column density of log \nhi=20.60. Akerman et al. (2005) have follow-up this system with high-resolution UV spectroscopy with the UVES instrument at VLT, and could derive the metallicity of absorber. They find [Zn/H]$\leq-$0.22 and [Cr/H]$\leq-$0.81 at \zabs=2.19289. Curran et al. (2005) have looked for 21-cm absorption in the background spectrum but could not detect any signal. Curran et al. (2008) have also looked for but not detected absorption due to OH molecule at radio wavelengths. This system is not seen in our SINFONI data, although a possible broad emission line is detected 200 km/s away from the expected position of the H-$\alpha$ line.

\subsection{Q1323$-$0021, \zabs=0.716}

This SDSS quasar was first observed at somewhat higher-resolution by Khare et al. (2004) with the Multiple Mirror Telescope (MMT). These authors derive log \nhi=20.21$^{+0.21}_{-0.18}$ from Voigt profile fitting of the
damped Ly-$\alpha$ line in the publicly available HST/STIS spectrum (PI: Rao). They have also reported a super-solar metallicity. 
The metallicity of this system was further studied by P\'eroux et al. (2006a) based on a VLT/UVES spectrum. 
These authors have used a complex velocity profile fit
with a total of 16 components extended over 215 km/s and deduced the following abundances with respect to solar: [Mg/H]$>-$0.58; 
[Fe/H]=$-$0.51$\pm$0.20; [Zn/H]=+0.61$\pm$0.20; [Cr/H]$<-$0.52; [Mn/H]=$-$0.37$\pm$0.20 and
[Ti/H]=$-$0.61$\pm$0.22. While this system is one of the most metal-rich absorbers known to date, it remains undetected in our SINFONI data. 

Interestingly, Hewett \& Wild (2007) and Chun et al. (2010) report a detection 1.25"-away from the quasar from broad-band K-band imaging from UKIRT and K-band from Keck observations, respectively. Straka et al. (2010) did not detect this object in lower resolution optical and near-IR images from APO telescope. While such observations are more sensitive to the continuum of galaxies, SINFONI data best detect emission lines. Therefore, these different findings are not inconsistent with each other, but indicate a low-SFR object ($<$0.1M$_{\odot}$/yr). All these pieces of information combined (high abundance, $J$/$R$ colours higher than expected for early types) seem to suggest that the object is an early type galaxy in line with expectations from Khare et al. (2007), although such objects usually contain little gas.

\section{Results}

\subsection{Detection Rate}

\begin{table*}
\begin{center}
\caption{Properties of the primary targets detected in the SINFONI cubes.}
\label{t:detections}
\begin{tabular}{cccccccccc}
\hline\hline
Quasar 		  &$z_{abs}$  &$\Delta v^a$ &$\Delta$ RA  &$\Delta$ Dec &projected dist &physical dist &$\nii \lambda$6585/H-$\alpha$	&12 + log (O/H)		  \\
			  &		  &[km/s]	     &["]		   &["]			  & ["] 		   &[kpc] 			& &	 	 \\
\hline
Q0302$-$223{\bf $^b$}		&1.009	&1	     	&$-$0.50   &$+$3.12  &3.16	&25	&$<0.34$ 	&$<$8.6\\
Q1009$-$0026	{\bf $^c$}    	&0.887	&125    	&$+$5.01  &$+$0.37  &5.01	&39	&$0.48^{+0.56}_{-0.27}$ &8.7$\pm$0.2\\
\hline\hline 				       			 	 
\end{tabular}			       			 	 
\end{center}			       			 	 
\begin{minipage}{140mm}
{\bf $^a$:} Observed velocity shift between the \zabs\ and \zem\ of the detected galaxies. \\
{\bf $^b$:} The higher-resolution HST/WFPC2 data from Le Brun et al. (1997) clearly shows that the object is subdivided into two sub-components, consistent with the elongated shape seen in the SINFONI data presented here. In this table, however, the object is treated as only one. The \nii\ $\lambda$ 6585 lines is badly contaminated by OH sky lines blended with \nii\ and only an upper limit could be measured.\\
{\bf $^c$:} For Q1009$-$0026, there is also an OH line but it is much narrower than the line detected.\\
\end{minipage}
\end{table*}			       			 	 

In summary, in the study presented here, we have detected 2 galaxies out of the 6 intervening DLAs and sub-DLAs searched for in just a couple of hours of observing time per target. With the caveat that the statistics of this sample are obviously small, the detection rate is 30\%. It is interesting to note that a Science Verification programme with SINFONI also aiming at studying DLA galaxies, detected one out of two targets (Bouch\'e et al., in preparation). These results clearly demonstrate the benefits of using IFU (Integral Field Unit) techniques to look for quasar absorbers in emission. This makes the approach presented here the most efficient ground-based search for DLA galaxies to date. It is noted, however, that in our target selection we have specifically selected metal-rich absorbers ([Zn/H]$>-$1.0 or upper limits of that order) with the hope to increase the likelihood of finding strong emitting galaxies (M\o ller et al. 2002). However, it can well be that some of the targeted galaxies are bright but not strong emitters (as noted for the sub-DLA towards Q1323$-$0021). Indeed, 25\%\ of Lyman Break Galaxies (LBGs) do not have Ly-$\alpha$ emission lines. Better statistics are needed to see if the absorber abundance is a key factor in allowing detection of absorber galaxies directly. If so, then deeper imaging of some current non-detections might be worthwhile. Overall, in the present work, the detection rate of H-$\alpha$ emission for intervening galaxy absorbers with metallicity around or greater than 1/10$^{th}$ solar is two out of six, one with super-solar absorption metallicity and one with [Zn/H]=$-$0.51$\pm$0.12.

Similarly, larger samples are necessary to assess the evolution of the detection rates of SINFONI surveys with redshift. For \mgii\ absorbers with equivalent widths $>$2 \AA, Bouch\'e et al. detect 14 absorbers out of 21 searched for (67\%) at z$\sim$1 (Bouch\'e et al. 2007). However, at z$\sim$2, only 2 \mgii\ out of 19 searched for (10\%) were found (Bouch\'e et al., private communication). In the sample presented here, both detected galaxies correspond to absorbers at z$\sim$1, while the absorber at z$\sim$2 remains undetected maybe because there are less metal-rich than their low-redshift counterparts.

Table~\ref{t:detections} summarises several properties of the two detections from the present study: velocity shift between  \zabs\ and \zem, impact parameter to the quasar line-of-sight in arcsec and kpc and rough estimates of the size of the detected objects in kpc. One can see that the impact parameters probed can be quite small. 

\subsection{Comparison with Other Detected Absorbers}

\begin{figure*}
\begin{center}
\includegraphics[height=23.5cm, width=16.cm, angle=0]{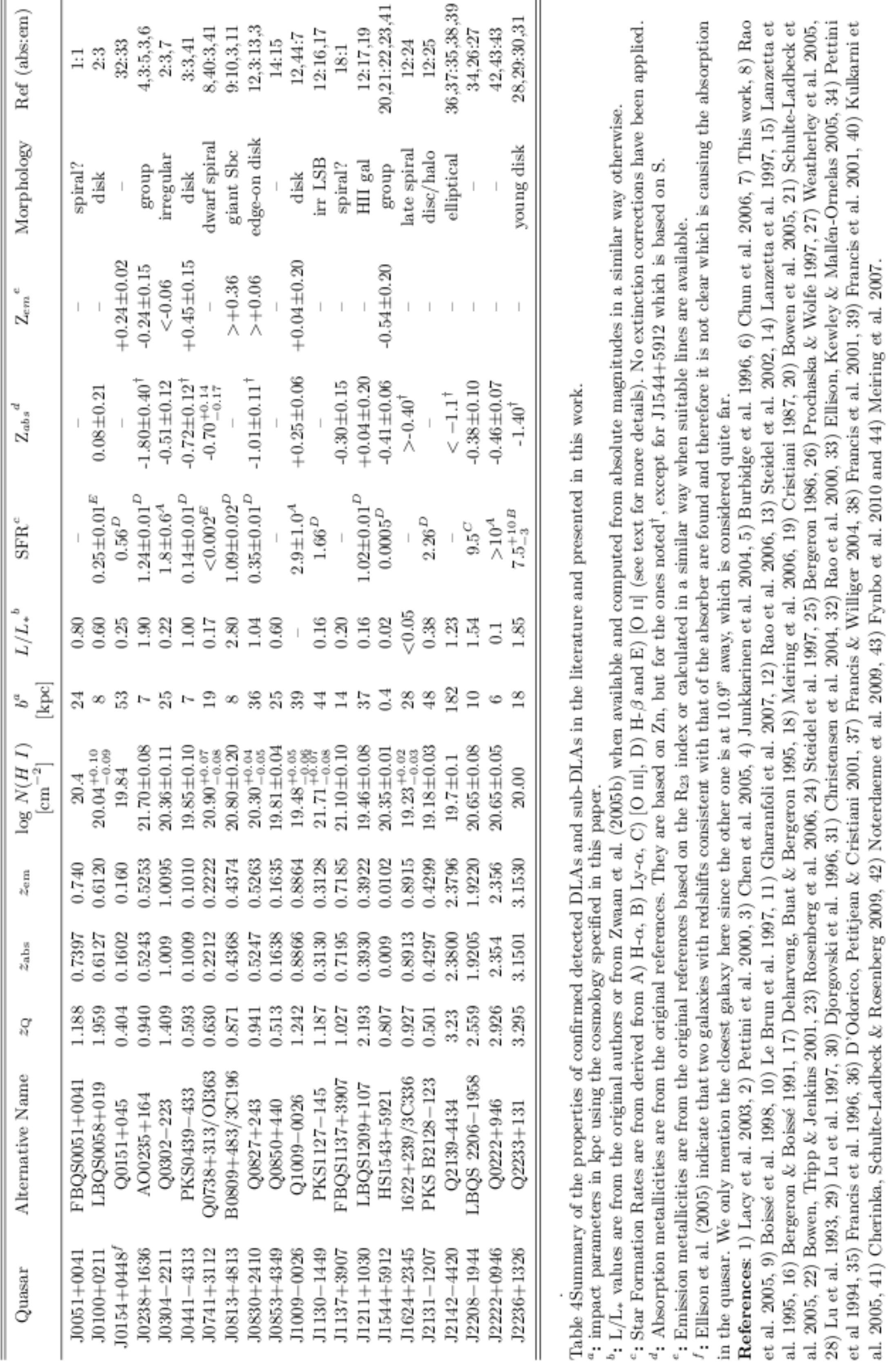}
\end{center}
\end{figure*}

\begin{figure*}
\begin{center}
\includegraphics[height=12cm, width=19cm, angle=0]{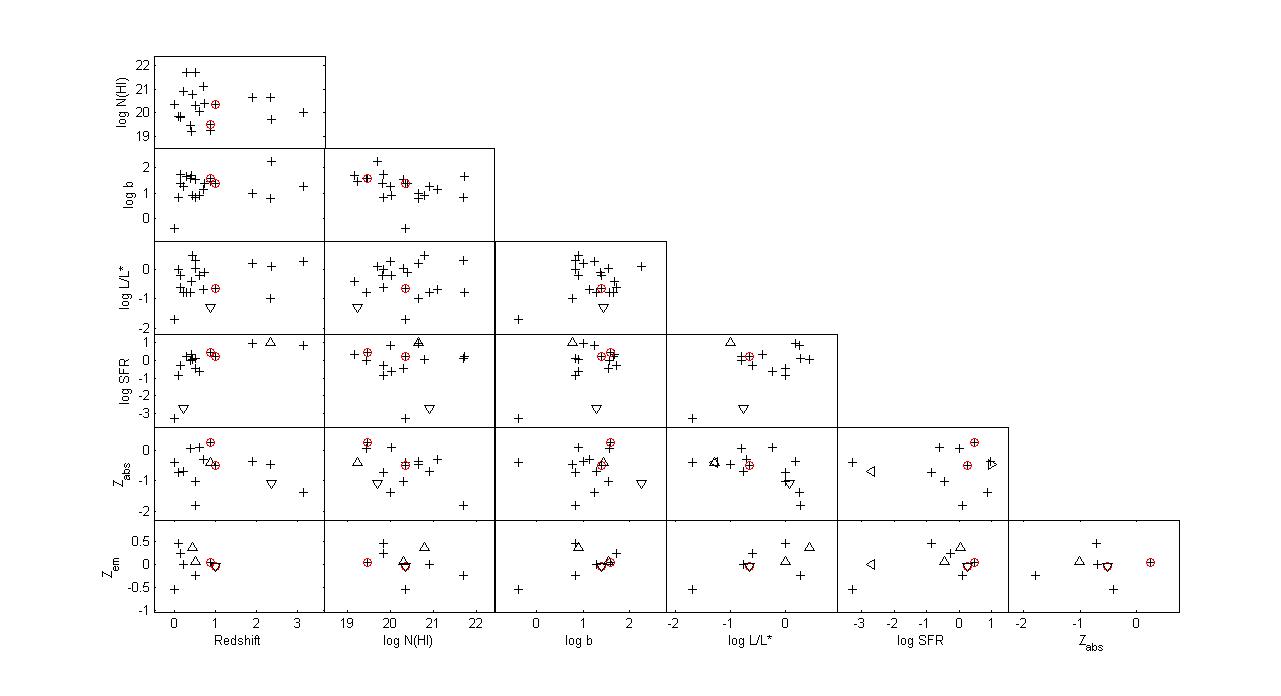}
\caption{Plots of the quantities listed in Table 4: redshift, \nhi\ column density, impact parameter in kpc, luminosity, star formation rate in solar masses per year, metallicity with respect to solar measured in absorption, Z$_{abs}$, (\hi\ abundance) and metallicity with respect to solar measured in emission, Z$_{em}$, (HII abundance). The triangles represent limits and the measurements reported in the present paper are circled in red.}
\label{f:Table5}
\end{center}
\end{figure*}

We have made a careful reappraisal of published reports of detections of galaxies responsible for DLA and sub-DLAs (see Table 4). We have only included systems with \nhi$>$19.0 (P\'eroux et al. 2003a), which are not associated with the quasar, for which \nhi\ was known from UV spectroscopy and for which the redshift of the emitting galaxy was confirmed spectroscopically. Most of these searches rely on the detection of the Lyman-$\alpha$ emission lines in imaging with subsequent spectroscopic confirmation. Note that the objects reported in Christensen et al. (2007) are not included in this table given the current limited quality of the spectra available for the confirmation of the redshift of the objects. The table does not include narrow-band imaging detections either, because there can always be confusion with other redshift interlopers.
This table does not include cases where Ly-$\alpha$ emission is detected in the trough of the Damped Lyman-$\alpha$ absorber because no information on impact parameters are then available. It does not include Gamma Ray Burst (GRB)-DLAs since most are associated with the GRB itself and since the fading of the burst precludes from studying the absorption properties at high-resolution (\nhi, metallicity). Our table, however, includes systems for which the galaxy was known before the absorber was found. These are the systems towards J1544$+$5912, J1211$+$1030, J0154$+$0448 and J2142$-$4420. We have increased the sample by including \mgii\ systems which were imaged in the 1990s (e.g. Bergeron et al., Steidel et al.) for which the \nhi\ has since then been measured from HST spectra (see Rao, Nestor \& Turnshek 2006 for a compilation). This makes a secure sample of identified quasar absorbers, but does not preclude that the galaxy imaged is part of a group with additional fainter galaxies at similar redshifts. We consider unlikely, however, that the identification is a mismatch given the proximity in the absorber and emission redshifts and the small impact parameters involved.

In Table 4, we include luminosity L/L$_*$ for each of the systems for which the continuum was detected in imaging. The values are taken from the original authors or from Zwaan et al. (2005) when available. We completed the list by estimating the absolute magnitudes from published apparent R or B magnitudes mostly, and on one occasion from K (J2142$-$4420). For the values of L$_*$, we adopted the measurement of M$^*_B$ - 5 log h = -20.3 from Norberg et al. (2002), M$^*_R$ - 5 log h = -21.1 from Lin et al. (1996) and M$^*_K$ - 5 log h = -24.1 from Cole et al. (2001). The Star Formation Rates (SFR) were derived from H-$\alpha$ in the case of the two absorbers presented in this study and for J2222$+$0946 studied by Fynbo et al. 2010. Indeed, Moustakas et al. (2006) argue that H-$\alpha$ luminosity is a reliable SFR tracer. In one case (J2236$+$1326), we used the estimate from the original reference (Djorgovski et al. 1996) based on Ly-$\alpha$ and in another occasion (J2208$-$1944), we used the estimate provided based on \oiii\ (Weatherley et al. 2005). Otherwise we used preferentially H-$\beta$ when available, assuming H-$\alpha$/H-$\beta$=2.85 (Osterbrock 1989) or on two occasions \oii\ (for J0100$+$0211 and J0741$+$3112), assuming \oii/H-$\alpha$=0.68 (Moustakas et al. 2006). In all cases, the estimates are not corrected for dust extinction in the galaxy nor by the Galaxy.

Figure~\ref{f:Table5} shows the relation between several of the properties listed in Table 4. The triangles represent limits and the measurements reported in the present paper are circled in red. From this figure, it appears that the sample of detected objects are mostly low luminosity and low SFR galaxies (Kulkarni et al. 2006), yet the galaxies have a wide range of luminosities in line with the morphologies collated in Table 4. We also note that small impact parameters (b$<$30 kpc) dominate and that very few galaxies are identified beyond 40 kpc which matches well the search radius in our SINFONI study at z$\sim$1. Taking out the outlier J1130$-$1449 with log \nhi=21.71 and b=44 kpc, we find a weak negative correlation where higher \nhi\ have smaller impact parameters (Spearman's rank correlation coefficient $r_s$=$-$0.42). Also it is interesting to note that the highest \nhi\ column density seem to have higher L/L$_*$ ratio. Indeed, the Spearman's rank correlation coefficient ($r_s$=0.30) indicates that the two are weakly correlated. 
Although the scatter is large, the reported absorption metallicities do not vary much with impact parameter. It is also clear that abundances reported in absorption are not well-correlated with emission abundances (see next paragraph for more discussion on this point).

\subsection{Metallicity}

\begin{figure}
\begin{center}
\includegraphics[height=8cm, width=10cm, angle=0]{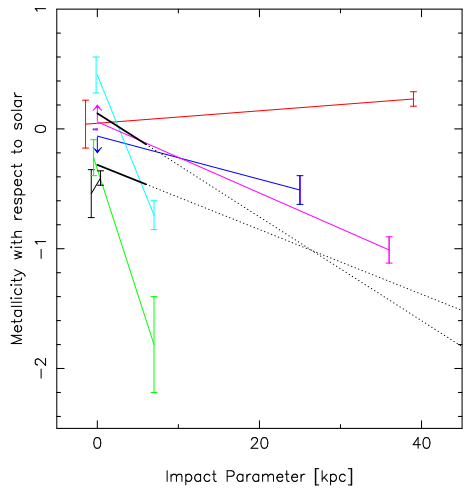}
\caption{Metallicity with respect to solar measured in emission at impact parameters b=0 and in absorption at given impact parameters in kpc. The two metallicities arising from the same galaxy are linked by coloured lines. The arrows indicate upper and lower limits on the measure of the emission metallicities for J0304$-$221 and J0830$+$2410, respectively. The errors on the emission metallicities are artificially offset from b=0 for clarity. The two solid lines are the HII region oxygen abundance measured in M101 (slope=$-$0.043 dex/kpc; Kennicutt, Bresolin \& Garnett 2003) and M33 (slope=$-$0.027 dex/kpc; Rosolwsky \& Simon 2008) for comparison.}
\label{f:Gradients}
\end{center}
\end{figure}

\begin{figure}
\begin{center}
\includegraphics[height=7cm, width=8cm, angle=-90]{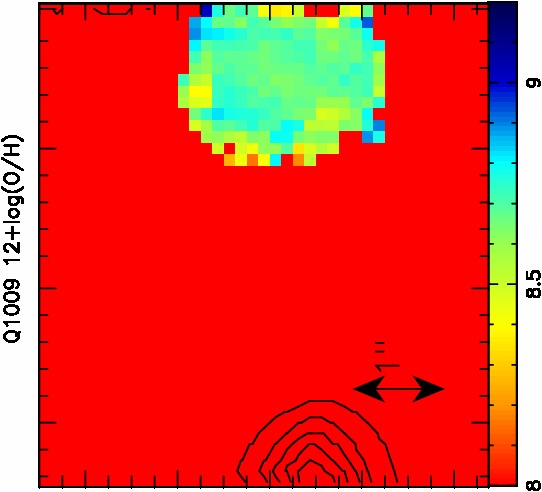}
\caption{This figure presents the metallicity in units of 12+log(O/H) derived from the N2 parameter (Pettini \& Pagel 2004), i.e. the ratio of \nii\ $\lambda$ 6585.27/H-$\alpha$, in Q1009$-$0026. The metallicity appears to be rather uniform on that scale as expected from the magnitude of the gradients observed in such objects (see Figure~\ref{f:Gradients}).}
\label{f:Q1009_metal_map}
\end{center}
\end{figure}

Table 4 includes metallicities measured in the neutral gas phase from absorption studies when these are available. The values are taken directly from the original references. Because Zinc is believed to be weakly depleted onto dust-grains, we preferentially report dust-free [Zn/H] metallicities when these are available and [Fe/H] otherwise. On one occasion (J1544$+$5912), the metallicity is derived from [S/H]. We note that there is a limited number of systems with dust-free estimates and only very few of these also have metallicity measured in emission. 

We have also included estimates of the HII metallicities derived from emission lines when sufficient information is available. The estimates are based on the R$_{23}$ index (Pagel et al. 1979) that uses the ratio of \oii+\oiii\ lines to H-$\beta$ to estimate 12+log(O/H), except for the two absorbers presented in this work, where the metallicities are estimated from the N2 parameter (see below). The R$_{23}$ indicator is double valued, but we consider the upper branch estimates since otherwise the galaxies will fall off the luminosity-metallicity relation for star-forming galaxies. Again, the values reported are from the original references except for J1211$+$1030 (Cristiani 1987), where we have calculated R$_{23}$ index using the published line intensities and the Kewley \& Dopita (2002) R$_{23}$ diagnostic diagram.

Unlike most of the other objects in Table 4 and other objects detected at these redshifts, the galaxies that we have been detected in this study have well-known absorption metallicities from high-resolution UV spectroscopy. Using the N2-parameter (Pettini \& Pagel 2004) based on \nii $\lambda$ 6585/H-$\alpha$ ratio, we can derive an estimate of the emission metallicity. For the DLA towards Q0302$-$223, we find 12+log(O/H)$<$8.6, i.e. less than solar\footnote{the solar abundance is 12+log(O/H)$_{\odot}$=8.66 (Asplund et al. 2004).} compared to the one-third solar absorption metallicity reported in absorption ([Zn/H]=$-$0.51$\pm$0.12). For the sub-DLA towards Q1009$-$0026, we find 12+log(O/H)$=$8.7$\pm$0.2, i.e. slightly above solar compared with a super-solar absorption metallicity ([Zn/H]=+0.25$\pm$0.06). These values are summarised in Table~\ref{t:detections}.

To investigate the role that gradients might play in the difference between the emission and absorption abundances, we have plotted the metallicity with respect to solar as a function of impact parameters in Figure~\ref{f:Gradients}. The two metallicities arising from the same galaxy are linked by dotted lines. Negative metallicity gradients are common in local spirals. For example, Rosolowsky \& Simon (2008) report a slope of $-$0.027 dex/kpc towards M33 from measurements of 61 HII regions, while Kennicutt, Bresolin \& Garnett (2003) find a slope of $-$0.043 dex/kpc for M101. However, it is interesting to note that at z$\sim$3, there has been detections of inverted gradients in about 30\% of the Lyman-$\alpha$ break-like galaxies, but on a much smaller distance scale (Cresci, private communication). 

Clearly, in most of 6 cases plotted here, the metallicity measured at a radius is lower than the one measured in the center, sometimes with a slope steeper that what is observed in local galaxies. In the case of Q1009$-$0026, however, the metallicity at the center of the galaxy
(seen in emission) is lower than the metallicity at 39 kpc radius (seen in
absorption). This surprising result might just be the result of an uncertain measurement as reflected by the large error bar (see Figure~\ref{f:Gradients}). It might also be partly explained by the use of the metallicity indicator. Indeed, Pettini \& Pagel have demonstrated that the N2-parameter/metallicity relation has a rather large scatter at high metallicities, where the data presented here lie. This illustrates the need to detect other emission lines in these objects in order to better constrain their metallicities using parameters such as the R$_{32}$ index. In addition, it should be emphasized that these measurements are not tracing the same gas phase (neutral versus ionised gas). It would be interesting to further increase the number of objects with metallicities both in absorption and in emission to better understand the reported low metallicities in DLAs (e.g. Kulkarni et al. 2007).

Finally, we use the spatial information made available to us thanks to the 3D spectroscopy to compute a metallicity map of Q1009$-$0026. Figure~\ref{f:Q1009_metal_map} presents the metallicity in units of 12+log(O/H) derived from the N2 parameter (Pettini \& Pagel 2004), i.e. the ratio of \nii\ $\lambda$ 6585.27/H-$\alpha$. The metallicity gradient appears to be rather uniform on that scale as expected from the magnitude of the gradients observed in such objects (see Figure~\ref{f:Gradients}).

\section{Results for Non-Prime Absorption Line Systems}

Thanks to the broad wavelength coverage of SINFONI, it is possible to look for the emission from additional absorbers that fortuitously fall along the line-of-sight to the quasar fields under study. We therefore made an 'a posteriori' check of other known absorbers in the quasar spectra and look for corresponding emission lines of H-$\alpha$, H-$\beta$ or \oiii\ in the SINFONI data cubes. Contrary to the primary targets however, these non-prime absorbers have {\it not} been selected to optimise the observing strategy (in terms of OH line contamination for example).

\subsection{Q0134$+$0051}

In addition to the primary target mentioned in Section 3, P\'eroux et al. (2006b) report a couple of other absorbers along the same line-of-sight. A \zabs=0.618 system is presented with strong metal lines of \feii\ $\lambda\lambda$ 2249, 2260, 2344, 2374,
2382 and \tiii\ $\lambda$ 3384, making this system a likely DLA or sub-DLA. Unfortunately, none of the predominant redshifted emission lines fall in the coverage of our SINFONI data. 

Similarly, an absorber at \zabs=1.272 with log \nhi\ in the range 18.9$-$19.1 is reported. This 
system shows very strong \civ\ and \mgii\ doublets,
\siii\ $\lambda$ 1526, \alii\ $\lambda$  1670, \mgi\ $\lambda$  2852, \feii\ $\lambda$ $\lambda$  1608, 2374, 2382,
2600. The emission lines of \oiii\ are covered by our SINFONI data in a region free from OH lines but none are detected. From the non-detection of \oiii\ and assuming 1 for the unknown ratio of L(H-$\alpha$)/L(\oiii) (Law et al. 2009), we derive L(H-$\alpha$)$<1.0 \times 10^{41}$ erg/s corresponding to SFR$<$0.4 M$_{\odot}$/yr.

Finally, another absorber is detected at \zabs = 1.449 with strong
\siiv\ doublet, \siii\ $\lambda$ 1526, \cii\ $\lambda$ 1334, \nii\ $\lambda$ 1370, \civ\ doublet and
\feii\ $\lambda$ 2600. Again, H-$\beta$ and OIII doublet are covered by our SINFONI data but none are detected, corresponding to L(H-$\alpha$)$3.2< \times 10^{41}$ erg/s corresponding to SFR$<$0.6 M$_{\odot}$/yr.

\subsection{Q0302$-$223}

\begin{figure*}
\begin{center}
\includegraphics[height=8cm, width=8cm, angle=0]{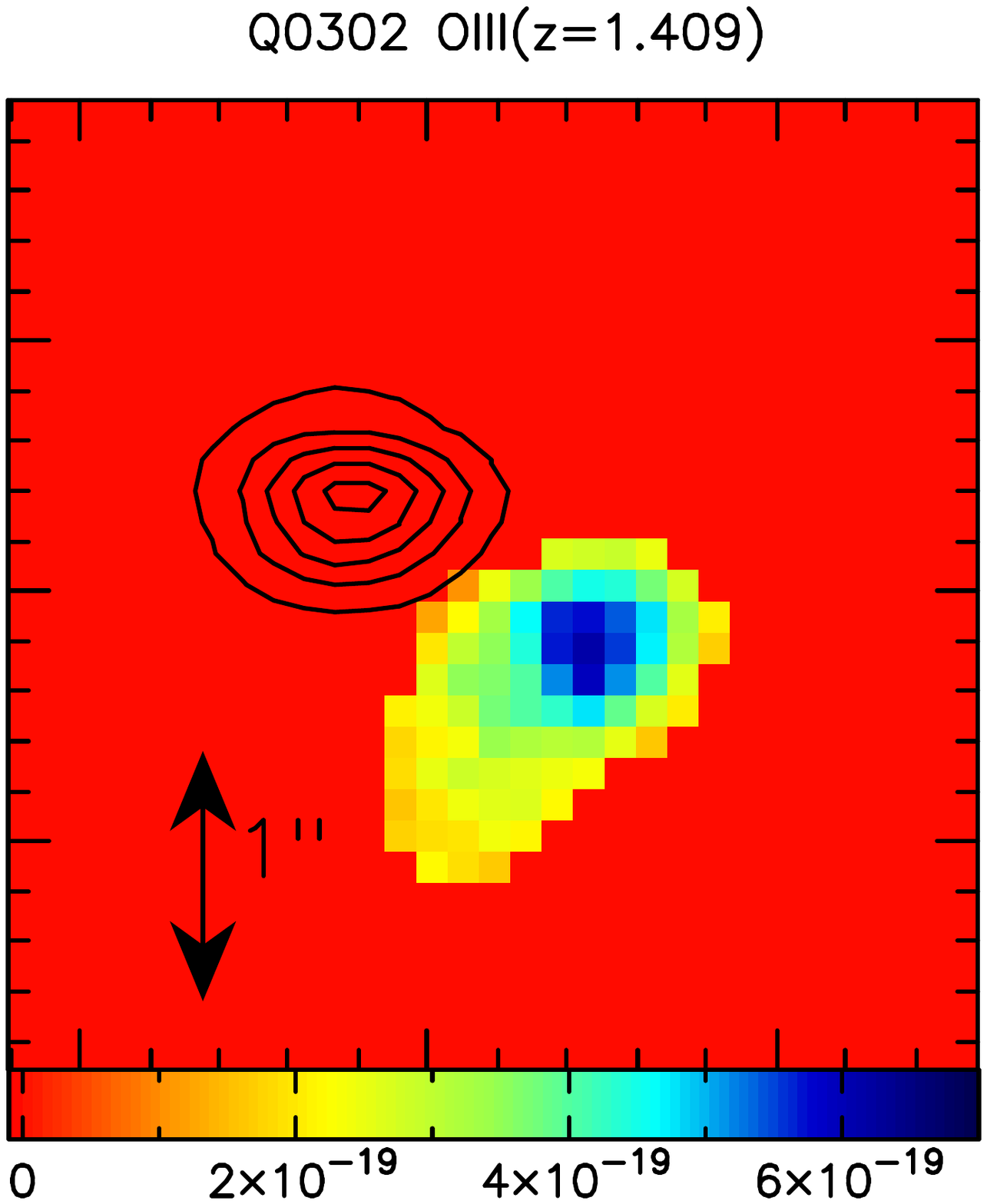}
\includegraphics[height=7cm, width=8.5cm, angle=0]{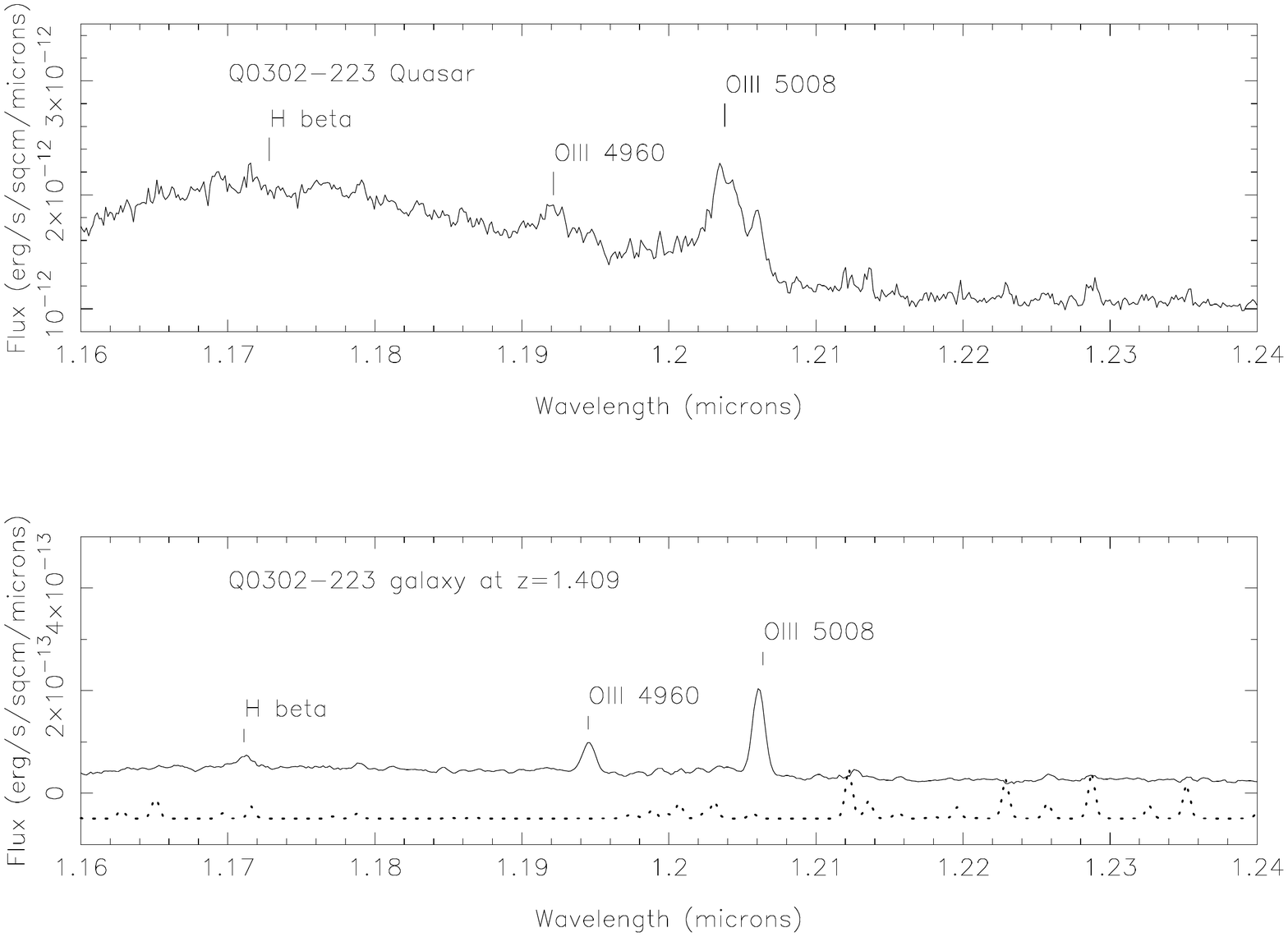}
\caption{{\bf Left Panel:} The \oiii\ emission map of the galaxy at the redshift of the quasar in the field of Q0302$-$223 (object \#2 in Figure~\ref{f:Q0302_HST}). This object is the closest in the HST image to the quasar, and was therefore mistakenly identified as the primary quasar absorber (Le Brun et al. 1997). The thin lined, black contours indicate the position of the quasar. At this redshift, 1"=8.5 kpc.{\bf Right Panel:} The integrated spectrum of the quasar (top) and that object (bottom). The emission lines of the object are clearly visible redwards of the quasar emission lines. The object spectrum is extracted from the SINFONI cube with a radius of 4 pixels to limit the contribution from the quasar in the vicinity. Nevertheless, the non-zero continuum limit is not intrinsic to that object, but due to the quasar itself. The spectrum is smoothed (5 pixel boxcar). Both the H-$\beta$ line and \oiii\ doublet are clearly detected. The dotted spectrum at the bottom of the panel is the sky spectrum indicating the position of the OH sky lines. }
\label{f:Q0302_Obj_spec}
\end{center}
\end{figure*}

As reported in Section 3, object \#2\ from Le Brun et al. (1997) is detected in with our SINFONI observations. This absorber is associated with the quasar and is also reported as a high-ionisation system with strong \ovi\ lines at \zabs=1.4055 by Boiss\'e et al. (1998). Both the H-$\beta$ line and \oiii\ doublet are clearly detected redwards of the quasar emission lines. The \oiii\ emission map and spectrum of this galaxy is shown in Figure~\ref{f:Q0302_Obj_spec}. Because of the proximity with the quasar, only a limit on the \oiii\ flux can be derived: it is measured to be F(\oiii)$>$16 $\times$10$^{-17}$ erg/s/cm$^2$. The corresponding luminosity of L(\oiii)$>$0.7 $\times$ 10$^{41}$ erg/s, leads to an estimated SFR of about $\sim$ 0.3 M$_{\odot}$/yr assuming 1 for the unknown ratio of L(H-$\alpha$)/L(\oiii) (Law et al. 2009). 

In addition to this target, the ultraviolet spectrum of this quasar 
exhibits strong candidate Lyman-$\alpha$ absorption features.
An absorber is reported at \zabs=0.118 (Bergeron, unpublished, Le Brun et al. 1997).
Boiss\'e et al. (1998) find relatively strong lines from \feii\ and \mgii. They argue that the relative strengths of \feii\ lines suggest
that \feii\ $\lambda$ 2374 and \feii\ $\lambda$ 2586 are nearly optically thin which rules out a value of log \nhi\
above 19. 
This absorber has no significant emission lines in our SINFONI coverage. 

However, Petitjean \& Bergeron (1990) report a strong \civ\ doublet together with Lyman-$\alpha$ at \zabs = 0.9109. The H-$\alpha$ line of this absorber is covered by our SINFONI data but not detected. Similarly, other absorbers are reported at \zabs=1.2482 (Boiss\'e et al. 1998) and a Lyman Limit System (LLS) at \zabs=1.3284 showing weak features
from \cii, \siii\ and \siiii\ (Koratkar et al. 1992, Boiss\'e et al. 1998). All of these have H-$\beta$ and/or \oiii\ doublet covered by our SINFONI data in a region free from OH lines, albeit in a part of the spectrum with strong atmospheric signatures, but remain undetected. The corresponding limiting luminosities are respectively: L(H-$\alpha$)$<0.4 \times 10^{41}$ erg/s, L(H-$\alpha$)$<0.9 \times 10^{41}$ erg/s
and L(H-$\alpha$)$<1.1 \times 10^{41}$ erg/s corresponding to SFR$<$0.2 M$_{\odot}$/yr, SFR$<$0.4 M$_{\odot}$/yr and SFR$<$0.5 M$_{\odot}$/yr.

\subsection{Q0405$-$331}

To our knowledge, no other absorber than the primary target have been reported in the literature along this line-of-sight. 

\subsection{Q1009$-$0026}

Besides the two primary targets mentioned in Section 3, no other absorbers have been reported along this line-of-sight.

\subsection{Q1228$-$113}

We have retrieved the VLT/UVES archive data for the quasar in order to check for any additional metal absorber that might be identified in the field. We detect a strong \civ\ absorber at \zabs=2.838. However, no strong emission lines fall into the available SINFONI coverage at this redshift. 

\subsection{Q1323$-$0021}

\begin{figure*}
\begin{center}
\includegraphics[height=7cm, width=8.5cm, angle=0]{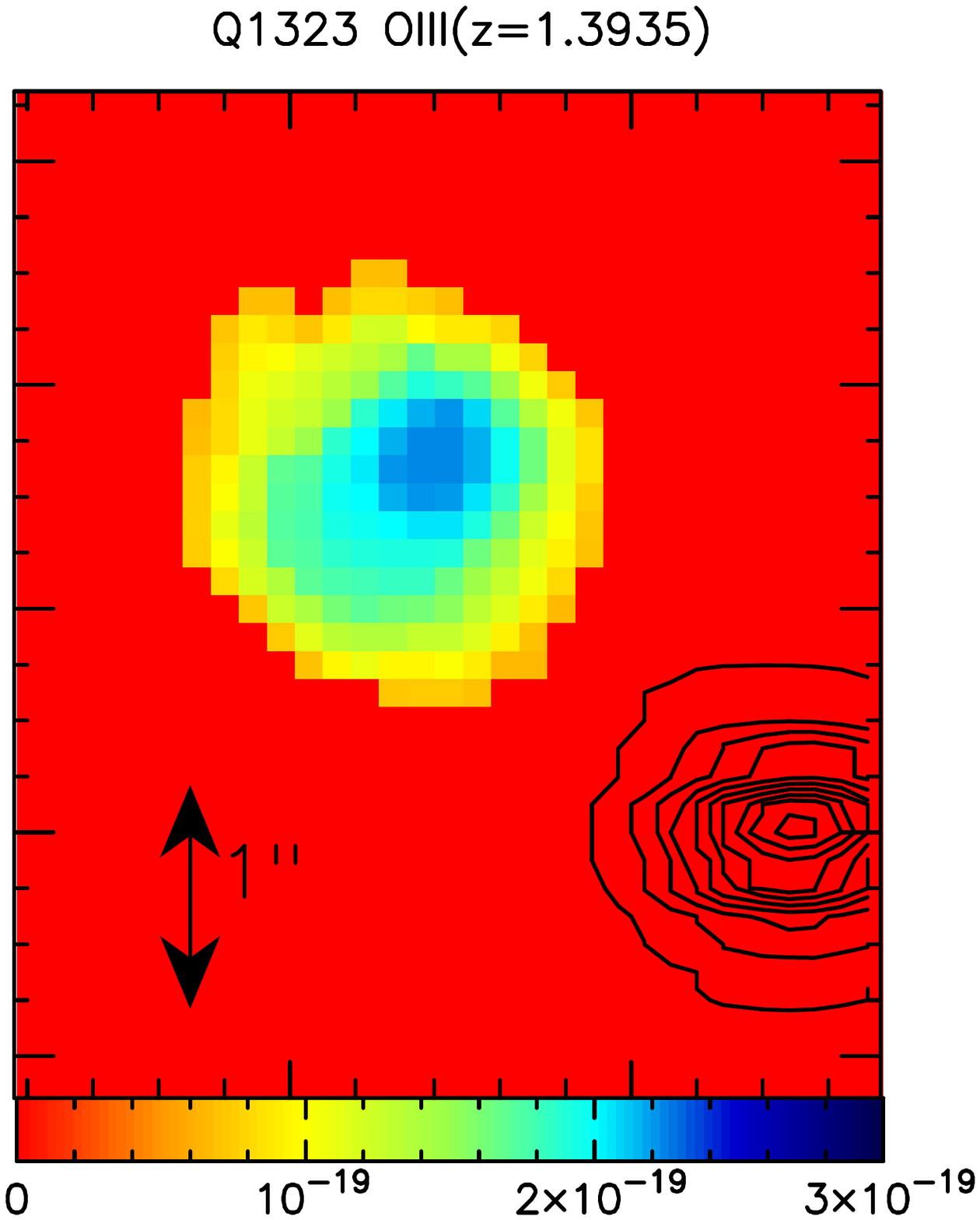}
\includegraphics[height=7cm, width=9cm, angle=0]{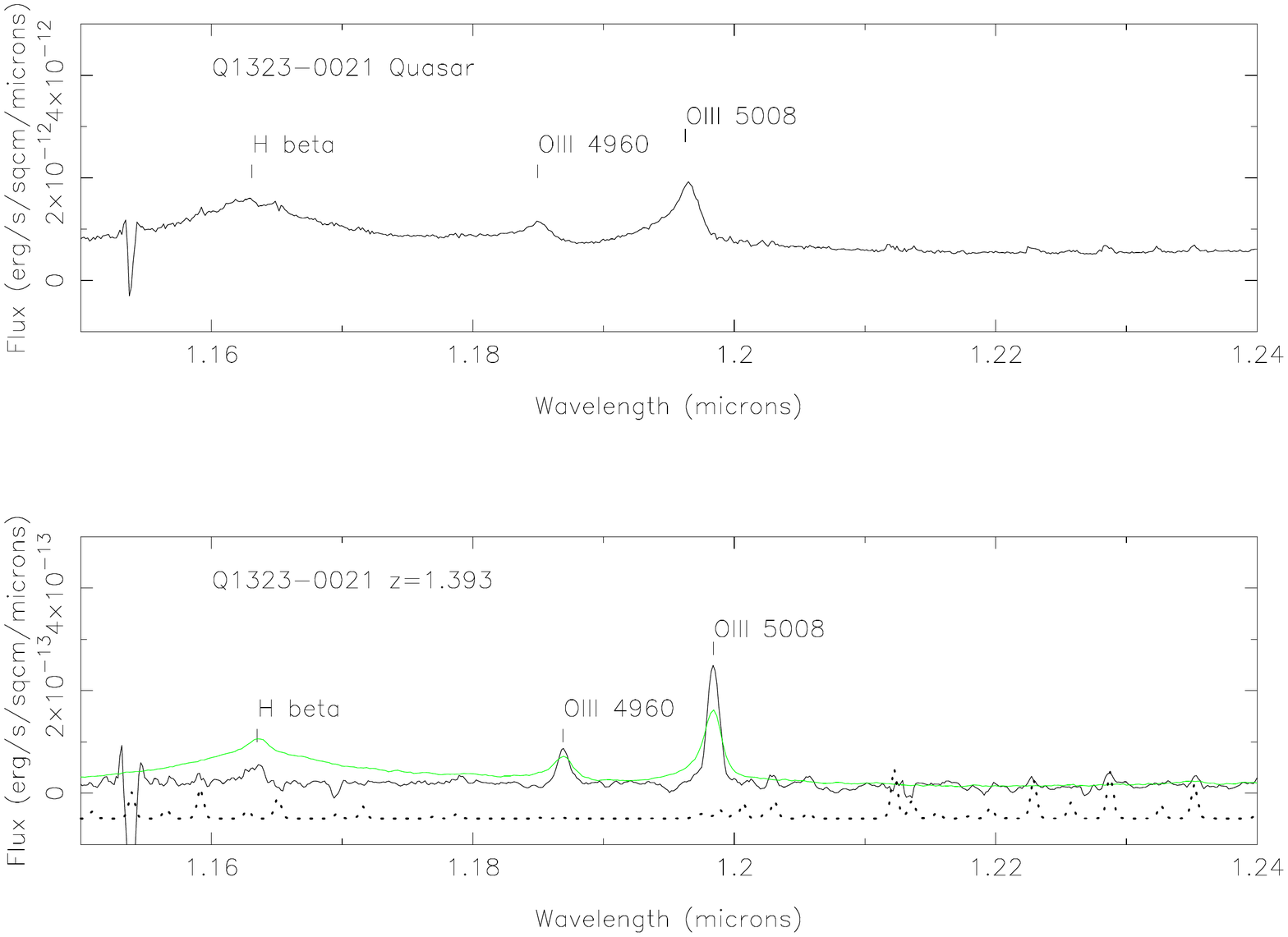}
\caption{{\bf Left Panel:} The \oiii\ emission map of an object serendipitously discovered in the SINFONI cube matching a strong \civ\ absorber complex (Khare et al. 2004, P\'eroux et al. 2006a) in the field of Q1323$-$0021. The object is slightly above the redshift of the quasar. The thin lined, black contours indicate the position of the quasar. At this redshift, 1"=8.5 kpc. {\bf Right Panel:} The integrated spectrum of the quasar (top) and the galaxy (bottom). Again, the emission lines of the galaxy are clearly shifted redwards of the quasar emission lines. The galaxy spectrum is smoothed (5 pixel boxcar). The H-$\beta$ and \oiii\ are clearly seen in this spectrum. The green line overplotted on the object spectrum is the SDSS quasar composite spectrum of Vanden Berk et al. (2001) redshifted to the object redshift and arbitrarily scaled to the observed flux. Clearly, the hydrogen line in the composite is much broader than in our object. The dotted spectrum at the bottom of the panel is the sky spectrum indicating the position of the OH sky lines. }
\label{f:Q1323_Obj_spec}
\end{center}
\end{figure*}

\begin{figure}
\begin{center}
\includegraphics[height=10cm, width=6cm, angle=-90]{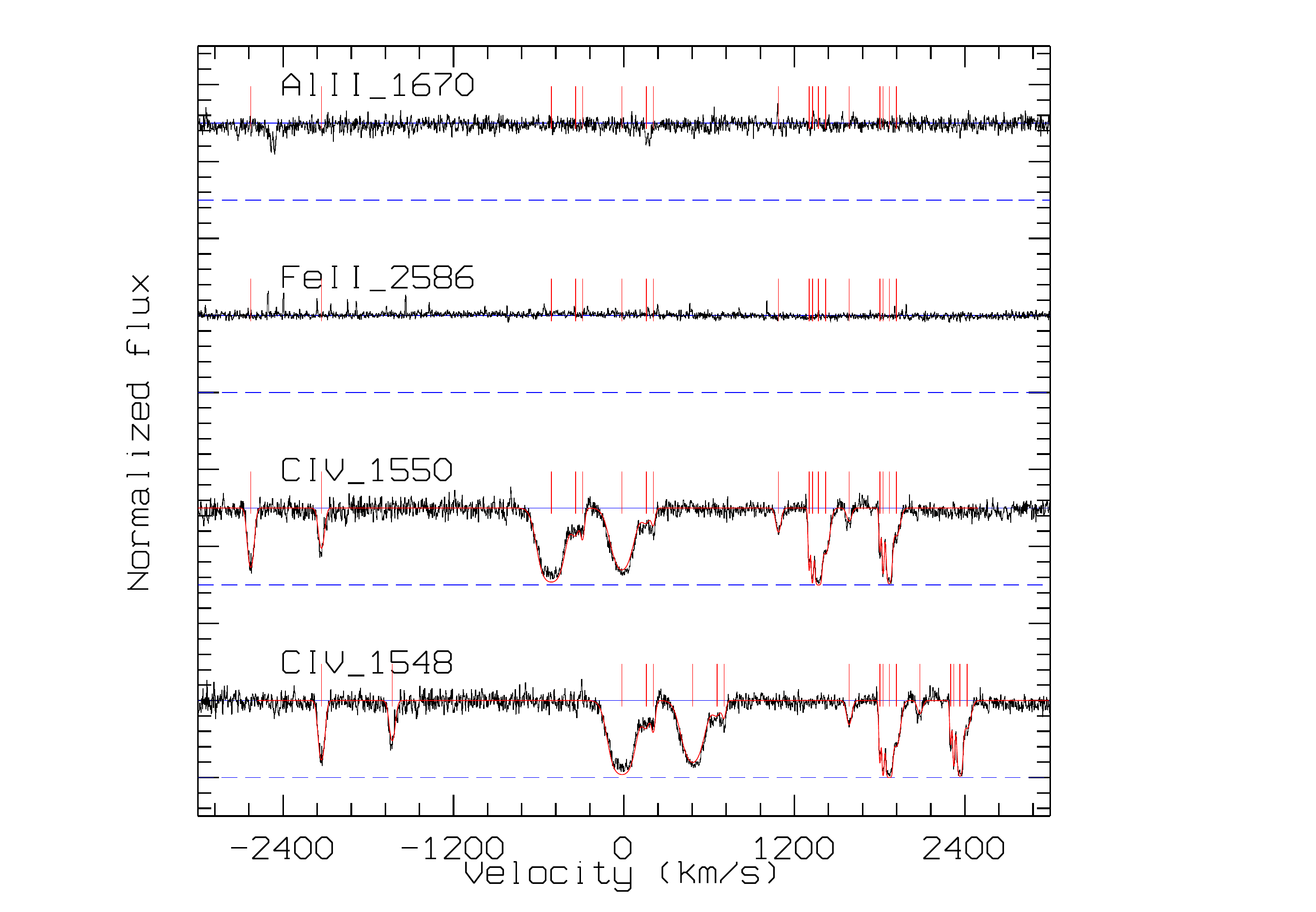}
\caption{Normalised UVES spectrum of Q1323$-$0021 showing the four \civ\ complexes along the line-of-sight to Q1323$-$0021 in velocity space. While the \civ\ lines are very strong in these systems, no other metal lines are detected. The velocity is set to \zabs=1.3745 for display purpose only, the absorption complex is $\Delta v$=390-4400 km/s from the galaxy seen in emission.
The red line in the \civ\ $\lambda$ 1548 and $\lambda$ 1550 panels is the Voigt profile fit to the systems and the red tick marks are the positions of the component fits for various species over the wavelength coverage presented. }
\label{f:Q1323_CIV}
\end{center}
\end{figure}

\begin{table}
\begin{center}
\caption{Parameters for the Voigt profile fit of the $z_{\rm abs} \sim 1.38$ \civ\ complex towards 
Q1323$-$0021 (see Figure~\ref{f:Q1323_CIV}). }
\label{t:Q1323_CIV}
\begin{tabular}{ccccc}
\hline\hline
         System$^a$      &$\Delta v$& $\lambda_{\rm {CIV}~1548}$    &N(\civ)   &b \\  
\hline
Syst B 	&$-$2119	&3650.16  &13.95$\pm$0.01  &26.3$\pm$0.7\\                  
Syst C 	&$-$14	&3676.02  &14.79$\pm$0.00  &83.6$\pm$0.8\\                  
       		&$+$157	&3678.12  &13.56$\pm$0.05  &40.0$\pm$3.9\\                  
       		&$+$207	&3678.73  &13.25$\pm$0.10  &17.3$\pm$2.8\\                  
Syst D 	&$+$1588&3695.69  &13.32$\pm$0.02  &24.1$\pm$1.4\\                  
Syst E 	&$+$1805&3698.36  &13.53$\pm$0.02  & 7.4$\pm$0.4\\                  
       		&$+$1828&3698.64  &13.91$\pm$0.01  & 8.0$\pm$0.3\\                  
       		&$+$1871&3699.17  &14.55$\pm$0.02  &18.5$\pm$0.6\\                  
       		&$+$1922&3699.80  &13.83$\pm$0.03  &32.9$\pm$1.9\\                 
\hline 				       			 	 
\hline 				       			 	 
\end{tabular}			       			 	 
\end{center}			       			 	 
\begin{minipage}{80mm}
{\bf Note:} The velocities are in km/s, the wavelengths are in \AA, column
densities are in log atoms/cm$^{2}$ and the doppler parameter values are in km/s.\\
{\bf $^a$:} Following Khare et al. (2004)'s nomenclature.
\end{minipage}
\end{table}			       			 	 

Khare et al. (2004) report a number of additional systems along this line-of-sight. The so-called systems B, C, D, and E are strong \civ-only systems at \zabs$\sim$1.38, all within 4000 km/s of each other. The higher-resolution VLT/UVES data clearly confirm all of the above systems as shown in Figure~\ref{f:Q1323_CIV}. We fit Voigt profiles to the \civ\ absorption lines to measure their column densities. The fits are performed using the $\chi^2$ minimisation routine {\tt fitlyman} in {\tt MIDAS} (Fontana \& Ballester 1995). The Doppler parameter $b$-value is left as a free parameter and the same redshift is used for both members of the doublet. The fit parameters are provided in Table~\ref{t:Q1323_CIV}. The resulting redshift estimates indicate that this complex is slightly above the redshift of the quasar. Interestingly, these absorbers have no other lines associated with them, not even \feii, \alii\ or \mgii\ and \mgi\ when these are covered. Indeed, the majority of the \civ\ systems in SDSS do not have associated \mgii\ (York et al., private communication). An inspection of the publicly available HST/STIS spectrum for this object does not indicate strong Lyman-$\alpha$ absorption either at the redshifts of the \civ\ absorbers. The absence of low-ionisation lines might indicate that this complex of systems is associated with the quasar host and hence substantially ionised as is common for the so-called associated systems in SDSS (Wild et al. 2008, Vanden Berk et al. 2008).  

The emission lines of H-$\beta$ and \oiii\ are seen at the position of \civ\ complex, redwards of the quasar emission lines (Figure~\ref{f:Q1323_Obj_spec}). Because of the proximity with the quasar, only a limit on the \oiii\ flux can be derived: it is measured to be F(\oiii)$>$27 $\times$10$^{-17}$ erg/s/cm$^2$. A spectrum of this object is also presented and shows very strong emission lines, suggesting that it might be associated with an active galactic nucleus (AGN). However, when comparing the width of the emission lines of the SINFONI spectrum with the SDSS composite quasar spectrum of Vanden Berk et al. (2001), the H-$\beta$ line seems much narrower than the average AGN in SDSS, and the \oiii\ lines considerably stronger (see the green spectrum overplotted in Figure~\ref{f:Q1323_Obj_spec}). 

We note again, that the galaxy at z$\sim$1.38 detected in the vicinity of the quasar line-of-sight is not at the redshift of the sub-DLA looked for (\zabs=0.716), and caution against possible over-interpretation based on broad-band imaging of quasar absorbers. The galaxy at z$\sim$1.38 is slightly above the redshift of the quasar and give rise to many absorbers spread over 4000 km/s maybe due to tidal interactions. Interestingly, the galaxy detected in the continuum by Hewett \& Wild (2007) and Chun et al. (2010) is on the same line from the quasar as the new galaxy seen in emission at z$\sim$1.38.

\subsection{Summary of the Non-Prime Absorption Line Systems}

In summary, in addition to the primary targets, we report the detections of two other objects in the field of Q0302$-$223 and Q1323$-$0021. Both of these objects are situated at redshifts similar to the quasar and very close to the quasar line-of-sight on the sky plane (impact parameters of 1.01" and 2.30", respectively). These have high-ionisation state metals, and therefore are perhaps probing the radiation field of the quasar to which they are proximate. 

This extended sample of non-prime absorbers provides a reference sample {\it not} selected on large \nhi\ nor on high metallicity (but not checked for possible OH line contamination at the wavelength of the expected emission lines a priori).
Such results are only possible thanks to the SINFONI set-up which allows to search for all galaxies in the field along the broad wavelength coverage. In addition, this observational approach provides an immediate redshift confirmation for the galaxy detected.

In the case of Q0302$-$223, the closest galaxy to the quasar's line-pf-sight was suggested as the quasar absorber candidate by Le Brun et al. (1997) based on the HST/WFPC2 observations. Thanks to the spectroscopic redshift confirmation made possible with the SINFONI data, we show that the DLA galaxy is in fact slightly further away in impact parameter. This galaxy was also detected in the HST/WFPC2 data where it shows a clear double structure. These results warn against rapid conclusion based solely on broad-band images, since in this particular case, it is not the object the closest to the line-of-sight which is responsible for the absorber. 

Table~\ref{t:detections_additional} summarises several properties of these two detections: \oiii\ fluxes, velocity shift between  \zabs\ and \zem, impact parameter to the quasar line-of-sight in arcsec and kpc and rough estimates of the size of the detected objects in kpc. 

\begin{table*}
\begin{center}
\caption{Properties of the non-prime objects detected in the SINFONI cubes.}
\label{t:detections_additional}
\begin{tabular}{cccccccccc}
\hline\hline
Quasar 		  &$z_{abs}$ &F(\oiii) &$\Delta v^a$ &$\beta^b$ &$\Delta$ RA  &$\Delta$ Dec &projected dist &physical dist 		  \\
			  &&[erg/s/cm$^2]$ 			  &[km/s]	&[km/s]     &["]		   &["]			  & ["] 		   &[kpc] 				 	 \\
\hline
Q0302$-$223{\bf $^c$}	&1.408	&$>$16 $\times$10$^{-17}$              &N/A     	&50 &$+$0.87  &$-$0.50   &1.01	&9	\\
Q1323$-$0021{\bf $^d$}	&1.35-1.39	&$>$27 $\times$10$^{-17}$               &390-4400 &380  &$-$1.62  &$+$1.62   &2.30	&20	\\
\hline\hline 				       			 	 
\end{tabular}			       			 	 
\end{center}			       			 	 
\begin{minipage}{140mm}
{\bf $^a$:} Observed velocity shift between the \zabs\ and \zem\ of the detected galaxies. In the case of the CIV systems along Q1323$-$0021, several systems are identified and therefore a range of $\Delta v$ is provided. \\
{\bf $^b$:} Observed velocity shift between the \zem\ of the detected galaxies and the \zem\ of the quasars.\\
 {\bf $^c$:} Object \#2 from Le Brun et al. (1997). In this case, the flux estimate is a lower limit because a small radius (4 pixels) was used to extract the spectrum from the SINFONI cube in order to avoid contamination from the nearby quasar continuum. \\
{\bf $^d$:} Again, the flux estimate is a lower limit because a small radius (8 pixels) was used to extract the spectrum from the SINFONI cube in order to avoid contamination from the nearby quasar emission lines in this case. \\
\end{minipage}
\end{table*}

\section{Conclusion}

We report the H-$\alpha$ detection of one DLA and one sub-DLA with well constrained \nhi\ and absorption metallicities, among a sample of six intervening quasar absorbers searched for. The use of integral field spectroscopy technique turns out to be an important ingredient of the high detection rate of H-$\alpha$ reported in this study. We derive F(H-$\alpha$)=7.7$\pm$2.7$\times$10$^{-17}$ erg/s/cm$^2$ (SFR=1.8$\pm$0.6 M$_{\odot}$/yr) at impact parameter b=25 kpc towards quasar Q0302$-$223 for the DLA and F(H-$\alpha$)=17.1$\pm$6.0$\times$10$^{-17}$ erg/s/cm$^2$ (SFR=2.9$\pm$1.0 M$_{\odot}$/yr) at b=39 kpc towards Q1009$-$0026 for the sub-DLA. We make a careful reappraisal of published reports of detections of quasar absorbers with log \nhi$>$19.0, and find 15 such detections at z$<$1 and only 3 such spectrospically confirmed detections at higher-redshifts. All of these objects are found to be faint, most of them having L/L$_*$$<$2 and when available, the SFR are measured to be few solar masses per year at most. Only a few of these systems have metallicity estimates in both absorption and emission, as is the case for new detections presented here. By finding a mis-identified absorbing galaxy in one of the fields under study, we show that spectroscopic follow-up of candidates from broad-band imaging studies are paramount for secure identifications. Further, results on the dynamical properties of these detections are presented in Paper II (P\'eroux et al. 2010). In conclusion, such studies as the one presented here should be enlarged to build a statistically significant sample of Damped and sub-Damped Lyman-$\alpha$ Systems to better constrain the properties of this population of galaxies as a whole.

\section*{Acknowledgements}
We would like to thank the Paranal and Garching staff at ESO for performing the observations in Service Mode. We thank Max Pettini for comments on an earlier version of the manuscript. CP thanks Benjamin Cl\'ement for helpful discussions on SINFONI data reduction and Thomas Ott for developing and distributing the QFitsView software. VPK acknowledges partial support from the U.S. National Science Foundation grants AST-0607739 and AST-0908890 (PI: Kulkarni). This work has benefited from support of the "Agence Nationale de la Recherche" with reference ANR-08-BLAN-0316-01.

\bsp

\label{lastpage}

\end{document}